\documentclass{WileyMSP-template}

\usepackage[separate-uncertainty=true]{siunitx}
\usepackage{graphicx}
\usepackage{dcolumn}
\usepackage{bm}
\usepackage{hyperref}
\usepackage[mathlines]{lineno}
\usepackage{xargs} 
\usepackage[pdftex,dvipsnames]{xcolor}  
\usepackage{amsmath,amssymb}
\usepackage{cite}

\newcommand{\chlamy}{\textit{C. reinhardtii}}
\newcommand{\chlamyfull}{\textit{Chlamydomonas reinhardtii}}
\newcommand{\cmax}{c_{\rm max}}
\newcommand{\uphoto}{\mathbf{u}_{\rm photo}}
\DeclareSIUnit[number-unit-product = {}]{\min}{min}
\DeclareSIUnit[quantity-product = {}]{\cells}{cells}
\newcommand{\ladhyx}{Laboratoire  d'Hydrodynamique (LadHyX), CNRS, Ecole Polytechnique, Institut Polytechnique de Paris, 91120 Palaiseau, France}
\newcommand{\fast}{Université Paris-Saclay, CNRS, FAST, 91405 Orsay, France}

\begin{document}

\pagestyle{fancy}
\rhead{\includegraphics[width=2.5cm]{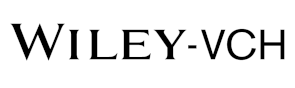}}

\title{Controlling the Collective Transport of Large Passive Particles with Suspensions of Microorganisms}

\maketitle


\author{Taha Laroussi}
\footnotemark{}
\author{Julien Bouvard*}
\footnotemark[\value{footnote}]
\footnotetext{These authors contributed equally to this work.}
\author{Etienne Jambon-Puillet}
\author{Mojtaba Jarrahi*}
\author{Gabriel Amselem}

\begin{affiliations}
Dr. T. Laroussi, Dr. J. Bouvard, Prof. E. Jambon-Puillet, Prof. G. Amselem\\
\ladhyx\\
Email Address: bouvard.julien@gmail.com

Prof. M. Jarrahi\\
\fast\\
Email Address: mojtaba.jarrahi@universite-paris-saclay.fr

\end{affiliations}


\keywords{Active matter, transport, bioconvection, microswimmers, decontamination, cargo delivery}

\justifying

\begin{abstract}

A promising approach to transport cargo at the microscale lies within the use of self-propelled microorganisms, whose motion entrains that of passive particles. However, most applications remain limited to just a few passive particles of similar size as the microorganisms, since the transport mechanism relies on the interaction between individual swimmers and single particles. Here, we demonstrate how to control the collective transport of hundreds of large passive particles with phototactic microalga. Using directional light stimuli in suspensions of \chlamyfull, we trigger bioconvection rolls capable of macroscale transport. Passive particles an order of magnitude larger than the microalgae are either attracted or repelled by the rolls depending on their density. Using experiments and simulation, we rationalize these bioconvective flows and describe how to harness them for cargo transport, with future applications in targeted drug delivery and decontamination.
\end{abstract}


\section{Introduction}


Despite several applications with high societal impact, such as soil and ocean remediation\textsuperscript{\cite{zarei2018self}} or targeted drug delivery,\textsuperscript{\cite{kei2014multiple,park2017multifunctional,mostaghaci2017bioadhesive}} controlled cargo transport remains a challenge at the submillimetric scale. These different applications have distinct goals and constraints, which prevents a ``one-size fits all'' solution. For instance, targeted drug delivery requires the precise transport of a relatively small cargo. For remediation purposes however, it may be more desirable to quickly sweep away pollutants, such as microplastics, from a large area to a collection point but without much precision.

Harnessing the motility of self-propelled microorganisms is a promising approach to control the transport of a passive cargo at the microscale. One method is to physically bind cargo, e.g. a bead, to the surface of motile cells, such as amoeba,\textsuperscript{\cite{nagel2019harnessing}} algae\textsuperscript{\cite{weibel2005microoxen}} or bacteria.\textsuperscript{\cite{fernandes2011enabling,kim2012chemotactic}} These beads are then carried by the cells as they crawl or swim. Control can be achieved by directing the cell motion with external cues to which cells respond, such as chemical gradients or light stimuli. Another strategy is to use the flows created by swimming microorganisms to displace suspended particles. These flows usually extend over a distance of roughly one microorganism ($\approx\qtyrange{1}{10}{\um}$), and can entrain neighboring colloids.\textsuperscript{\cite{jeanneret2016entrainment}} When confined inside chambers with specific branched geometries, the interaction between microalgae and passive particles can lead to demixing, with colloids eventually accumulating in certain regions of the device which could facilitate their collection for decontamination applications.\textsuperscript{\cite{williams2022confinement}}

However, all these transport mechanisms rely on the interaction between a single microorganism and a single passive particle, usually smaller than the microorganism itself. At larger scale, passive beads mixed in a suspension of active fluids such as swimming bacteria or microalgae move randomly, albeit with an effective diffusion coefficient several orders of magnitude larger than predictions from the traditional Stokes-Einstein relationship.\textsuperscript{\cite{wu2000particle,patteson2016particle,bouvard2023ostwald,leptos2009dynamics,kurtuldu2011enhancement,ortlieb2019statistics}} Increasing the concentration of passive particles, one can eventually observe segregation, with the formation of clusters ranging in size from as few as ten beads,\textsuperscript{\cite{gokhale2022dynamic,kushwaha2023phase}} to very large clusters up to $10^5$ particles.\textsuperscript{\cite{bouvard2023ostwald}} Yet, this large-scale clustering stems from random motions and cannot be easily controlled. To obtain a directional motion, one needs to break the system symmetry, which has previously been achieved by designing geometrically asymmetric structures that guide the swimmer's movement.\textsuperscript{\cite{di2010bacterial, sokolov2010swimming, koumakis2013targeted, kaiser2014transport, vizsnyiczai2017light, williams2022confinement, pellicciotta2023light}} However, this method requires careful manufacturing, and is not adaptable. The direction has to be planned in advance and a specific microdevice needs to be prepared.
Would it be possible to transport a large number of particles, without size constraints, and control the transport direction dynamically?

Here, we leverage collective effects in suspensions of the swimming microalgae \chlamyfull\ to achieve collective directional transport at the microscale. Rather than relying on surface patterning to direct the microswimmer's motion, we use a directional light stimulus which can be tuned dynamically. By locally accumulating microalgae which are denser than water, we trigger macroscale bioconvection patterns\textsuperscript{\cite{dervaux2017light,arrieta2017phototaxis,arrieta2019light,ramamonjy2022light,fragkopoulos2025metabolic}} that we harness to transport particles. Beads from $\qty{50}{\um}$ to $\qty{460}{\um}$, up to 50 times larger than a single microalga, are successfully transported collectively over several millimeters along a prescribed light path. We propose a simple continuous description of the algae suspension that we solve numerically to predict the flow profile and the bead motion, and find good agreement with experiments. Depending on the buoyancy of the beads, these bioconvection rolls can either direct particles to a target location or sweep them away from a given area, highlighting the versatility and potential of this system for applications.


\begin{figure*}[!ht]
    \centering    
    \includegraphics[width=1.0\textwidth]{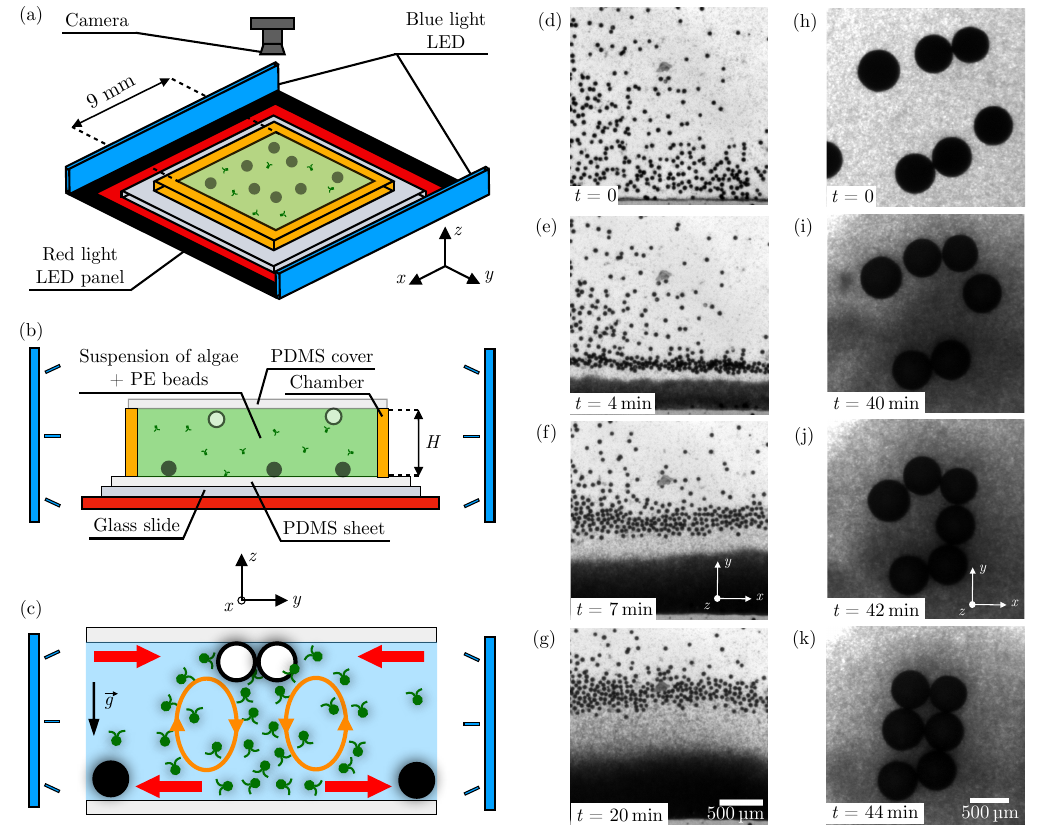} 
    \caption{\textbf{Experimental setup - Photo-bioconvective flows displace large beads.} (a) Schematic image of the setup. A square chamber of width $\qty{9}{\mm}$ and height $H$ is placed on top of a red light LED panel. The experiment is recorded with a camera placed above the chamber. Two blue light LED bands are placed on each side of the chamber. (b) Cross-sectional view of the chamber. The chamber of height $H$ is filled with a suspension of algae \chlamyfull\ and polyethylene (PE) beads. (c) Sketch of the recirculating convection rolls that appear when algae accumulate in a region of the chamber.
    Passive particles in range of those rolls are either attracted or repelled depending on their relative density with the medium. Dense beads are drawn in black and light beads in white. (d-g) Repulsion of denser beads of diameter $d_b=\qty{50}{\um}$, located on the bottom surface, in a chamber of height $H=\qty{310}{\um}$ (top view). As negatively phototactic algae accumulate at the lateral wall, beads denser than the fluid are pushed away, forming a front (see \href{https://seminaris.polytechnique.fr/share/s/wmNYoFtoTnya5FZ}{Supplementary~Movie~1}).
    (h-k) Attraction of lighter beads of diameter $d_b=\qty{460}{\um}$, located on the top surface, in a chamber of height $H=\qty{930}{\um}$ (top view). Both lateral LEDs are then switched on, forming a region of highly concentrated algae which draws in beads lighter than the fluid (see \href{https://seminaris.polytechnique.fr/share/s/aQJpTfqWfPjMjm3}{Supplementary~Movie~2}). Both experiments (d-k) are conducted with an initial optical density OD$_i$ of 10, i.e. $c_i=\qty{3e7}{\cells\per\mL}$.}
    \label{fig:fig1}
\end{figure*}

\section{Results}

\subsection{Beads move collectively in response to light stimuli}

Experiments begin by enclosing a suspension containing submillimetric polyethylene microspheres, of diameter $d_b=\qty{50}{\um}$ and median density $\rho_b \approx \qty{1006}{\kilo\gram\per\cubic\m}$ (see Methods), with microalgae \chlamyfull\ in a square rectangular chamber of side 9\,mm and height $H=\qty{310}{\um}$. The chamber is placed on top of a red LED panel ($\lambda = \qty{630}{\nm}$) for visualization purposes, and imaged from above. Two blue LED strips ($\lambda = \qty{470}{\nm}$) are positioned on opposite sides of the chamber to stimulate the algae, see sketches in \textbf{Figure~\ref{fig:fig1}}a,b and Materials and Methods for more details. Initially, the blue LEDs are off and the beads and algae are homogeneously distributed throughout the chamber, see Supplementary~Figure~1b. Unless noted otherwise, the algal suspension contains $c = \qty{3e7}{\cells\per\ml}$, corresponding to an optical density $\rm{OD} = 10$. 

When one LED strip is switched on, the algae respond by negative phototaxis and swim away from the intense blue light.\textsuperscript{\cite{laroussi2024short,ramamonjy2022nonlinear,eisenmann2025light}} This leads to all algae accumulating at the edge of the chamber opposite to the light source, see time-lapse in Figure~\ref{fig:fig1}d-g. At the same time, the microspheres suspended with the algae dramatically reorganize: while initially evenly distributed throughout the chamber, these beads, five times larger than an individual microalga, are collectively pushed away from the dense algal zone and form a front distinctly separated from the dense algal region, see time-lapse in Figure~\ref{fig:fig1}d-g and \href{https://seminaris.polytechnique.fr/share/s/wmNYoFtoTnya5FZ}{Supplementary~Movie~1}.

Repeating the exact same experiment with beads of density $\rho_b\approx \qty{990}{\kilo\gram\per\cubic\m}$ and diameter $d_b = \qty{460}{\um}$ leads to the opposite behavior: the zone dense in algae attracts nearby beads, leading to the formation of a raft of beads, see time-lapse in Figure~\ref{fig:fig1}h-k, \href{https://seminaris.polytechnique.fr/share/s/aQJpTfqWfPjMjm3}{Supplementary~Movie~2} and Supplementary~Figure~7. Note that the beads in this experiment are almost 50 times larger than a single microalgae. On a longer time scale ($\sim 1\,$h), these floating beads can `surf' over the concentrated algal regions, hopping on and off as they develop and disappear, see Supplementary~Figure~8.

Experiments with beads of density either $990$ or $\qty{1006}{\kilo\g\per\cubic\m}$ and diameters ranging between $d_b=50$ and $\qty{460}{\um}$ reveal that the collective motion of beads occurs systematically, for all bead diameters. Its direction depends solely on the bead density: dense beads are pushed away from the zone concentrated in algae, while buoyant beads are attracted towards it. This suggests the existence of a convective flow, going away from the dense region of algae at the bottom of the chamber and towards the algae at the top of the chamber, see sketch in Figure~\ref{fig:fig1}c. 

As shown in,\textsuperscript{\cite{dervaux2017light,arrieta2019light}} concentrating algae with light in localized regions can create buoyancy driven flows. Since algae are denser than water with a density $\rho_a \approx \qty{1050}{\kilo\g\per\cubic\m}$,\textsuperscript{\cite{dervaux2017light,arrieta2019light}} a lateral concentration gradient of algae therefore leads to a lateral density gradient (perpendicular to gravity). This situation is unstable,\textsuperscript{\cite{batchelor1954heat,guyon2012hydrodynamique}} and triggers a flow in which the denser region (algae-rich) falls and flows towards the lighter region (algae-poor) which itself rises and flows towards the denser region.
 
In the absence of forcing, this flow is transient since it mixes the fluid and thus erases the density gradient. However, in our case it is sustained as long as the light stimulus is on. Indeed, while algae are advected away from the dense region by the buoyancy driven flow, their negative phototaxis makes them swim back towards the dense region, away from the strong light. This repopulates the dense algal region\textsuperscript{\cite{dervaux2017light}} which reaches a steady-state, see sketch in Figure~\ref{fig:fig1}c. Note that, unlike the well-known Rayleigh-B\'{e}nard instability where temperature differences generate a density gradient parallel to gravity, this convective flow due to a lateral density gradient occurs without any threshold.\textsuperscript{\cite{batchelor1954heat,guyon2012hydrodynamique}}

The bioconvection rolls can be experimentally observed by following beads with a density $\approx \qty{1001}{\kilo\g\per\cubic\m}$ close to the surrounding medium's density. At first, these beads are pushed away from the region of high algal density. After traveling a few hundred microns, they enter upwelling flows, reaching the top lid before recirculating towards the side wall, eventually returning to the algae-dense region, see Supplementary~Figure~6a. Some beads can even stay inside the convection rolls for up to 1\,h, following the recirculating flows. Looking from the top, these beads seem to be `bouncing' on the edge of the dense algal region, see Supplementary~Figure~6b and \href{https://seminaris.polytechnique.fr/share/s/isWat6T4LsstFd6}{Supplementary~Movie~3}. Such bead trajectories highlight the simultaneous presence of inward (attractive) and outward (repulsive) flows at different heights within the chamber.

\subsection{A minimal numerical continuous model of photo-bioconvection yields quantitative predictions of bead trajectories}

To quantitatively understand these photo-bioconvection rolls and the transport of beads by the algae, we develop a continuous mathematical model of the suspension of algae. Bioconvection can be rationalized by idealizing algae as an active concentration field $c$ coupled to the Navier-Stokes equation describing the fluid velocity $\mathbf{u}$.\textsuperscript{\cite{pedley1992hydrodynamic}} Coarse-graining the microorganism microscopic motion results in an advection-diffusion equation for $c$ with an effective diffusion coefficient $D$ representing random motion, and a self-advection velocity representing directed motion due to any of the `taxis' (chemotaxis, gravitaxis, phototaxis, rheotaxis...). Coupling with the fluid flow occurs through the advection of $c$ but can also appear in the `taxis' (e.g. gyrotaxis or rheotaxis), through the mixture density, or in fluid stresses that can incorporate cell local stresses in addition to viscous stresses.

In an attempt to gain physical insights into the phototactic bioconvection rolls observed in our experiments, we keep the model as simple as possible. The experiment being invariant in the $x$ direction (see Figure~\ref{fig:fig1}), we only consider a 2D ($y$,$z$) slice of the chamber and incorporate only key physical ingredients: phototaxis with an advection velocity $\uphoto$ and density coupling $\rho(c)=\rho_w+\phi_a(\rho_a-\rho_w)$ with the Boussinesq approximation. Here, $\rho_w$ is the fluid density in the absence of algae, similar to the fluid density of water, $\rho_a$ is the density of algae, and $\phi_a=(4/3)\pi R_a^3c$ is the algal volume fraction, estimated assuming an individual alga is a sphere of radius $R_a\simeq\qty{5}{\um}$.\textsuperscript{\cite{arrieta2019light}} As previously done for phototactic \chlamy,\textsuperscript{\cite{dervaux2017light,arrieta2019light}} we thus neglect gravitaxis, gyrotaxis, cell stresses and assume all algae behave identically. The coupled equations for the concentration $c(y,z,t)$ and fluid velocity $\mathbf{u}(y,z,t)$ are:

\begin{align}
        \rho_w\left(\frac{\partial \mathbf{u}}{\partial t} + (\mathbf{u}\cdot\boldsymbol{\nabla})\mathbf{u}\right) &= -\boldsymbol{\nabla} p + \eta\Delta\mathbf{u} + \rho(c) \mathbf{g},\label{eq:NS1}\\
        \boldsymbol{\nabla}\cdot\mathbf{u}&=0,\label{eq:NS2}\\
    \frac{\partial c}{\partial t}+ \boldsymbol{\nabla}\cdot \left[(\mathbf{u}+\uphoto)c \right] &= D\Delta c.\label{eq:AD}
\end{align}
Here, $p(y,z,t)$ is the pressure, $\eta$ is the fluid viscosity and $\mathbf{g}$ is gravity.

We observe that algae swim away from the light and notice that they form a band of constant concentration $\cmax\approx\qty{3e8}{\cells\per\mL}$ at the lateral wall (see Figure~\ref{fig:fig1}e), corresponding to a volume fraction $\phi_{a,\mathrm{max}}<\qty{25}{\%}$, hinting that some physical processes that we have neglected prevent them from becoming more concentrated. These could include finite size interactions, local fluid stresses, or light shielding. For simplicity, we thus assume $\uphoto=-u_\mathrm{photo} (1-c/\cmax) \mathbf{e_y}$. This phototactic velocity simplifies to a constant velocity $-u_\mathrm{photo} \mathbf{e_y}$ in the dilute regime, while accounting for crowding effects when concentration increases.

We solve Equation~\eqref{eq:NS1}-\eqref{eq:AD} numerically in a rectangular domain of the experimental chamber dimensions ($\qty{9}{\mm}\times\qty{310}{\um}$) with the finite element solver COMSOL. We use a no flux boundary condition for the concentration $c$, a no slip boundary condition for the velocity $\mathbf{u}$, and assume a constant initial algae concentration $c(y,z,0)=c_i$ and zero initial velocity $\mathbf{u}(y,z,0)=\mathbf{0}$. In practice, we solve Equation~\eqref{eq:AD} for the rescaled concentration $\bar{c}=c/\cmax$ that varies between 0 and 1 and use $\rho(\bar{c})=\rho_w+\bar{c}(\rho_\mathrm{max}-\rho_w)$ in Equation~\eqref{eq:NS1} (see Methods). All model parameters apart from velocity $u_\mathrm{photo}$ can be measured experimentally. The diffusion coefficient $D=\qty{4e-9}{\meter\squared\per\second}$ is measured independently (see Supplementary~Section~I.C), $\rho_\mathrm{max}\approx\qty{1008}{\kilo\gram\per\cubic\meter}$ and $\rho_w\approx\qty{1000}{\kilo\gram\per\cubic\meter}$ are extracted from experimental images for each experiment, and $\eta=\qty{1e-3}{\Pa\second}$. The average swimming velocity of \chlamy\ is measured as \qtyrange{40}{60}{\um\per\s} under a microscope (see Supplementary~Figure~4). All phototactic speeds tested in the range $u_\mathrm{photo}\sim\qtyrange{1}{100} {\um\per\s}$ systematically lead to an accumulation of algae at the edge of the simulation box, and a self-sustained convective roll. Adding gravitaxis to account for the bottom heaviness of the algae\textsuperscript{\cite{bees2020advances}} or taking a crowding term of the form $(1-c/\cmax)^n$ with $n\neq 1$ has only a small influence on the steady-state convective roll characteristics (see Supplementary~Figure~9 and Section II.B. in Supplementary).

To be more quantitative, we focus on dense ($\rho_b \approx \qty{1006}{\kilo\g\per\cubic\m}$) beads of diameter $d_b = \qty{50}{\um}$ in algal suspension at concentration $c_i = \qty{3e7}{\cells\per\ml}$ (corresponding to $\mathrm{OD}_i=10$). A typical bead displacement tracking is shown in \textbf{Figure~\ref{fig:fig2}}a (see Supplementary Material for details). To compare experimental results to simulations, we measure the initial algae concentration $c_i(y)$ along the $y$ direction, before the blue-light is turned on, and use it as the initial condition in the simulations. We take an advection velocity $u_\mathrm{photo}=\qty{35}{\um\per\s}$, which matches the experimental time required for the algae to accumulate on one side of the chamber. We then obtain the density/concentration profile and fluid velocities shown in Figure~\ref{fig:fig2}b and \href{https://seminaris.polytechnique.fr/share/s/qbYKmyknoKSa9CG}{Supplementary~Movie~4}. Initially, the algae concentration is small and almost uniform. As the algae swim to the left and concentrate near the lateral wall, a weak convection roll immediately forms at the wall. This convection roll becomes stronger and wider as more algae accumulate, with fluid velocities reaching up to $||\mathbf{u}||\approx \qty{25}{\um\per\s}$. Eventually, the concentration at the wall reaches $c\approx \cmax$. After this point, more algae accumulation creates a growing band at $c\approx \cmax$ near the wall, and the concentration front between the algae-rich and poor regions moves away from the wall. The convection roll moves with the front while keeping the same velocity magnitude and width, see \href{https://seminaris.polytechnique.fr/share/s/qbYKmyknoKSa9CG}{Supplementary~Movie~4} for a more detailed view of the roll dynamics. The beads velocities and trajectories are calculated assuming they behave as passive tracers (see Methods) and shown alongside experiments in Figure~\ref{fig:fig2}c for three representative beads (dashed lines).

\begin{figure*}[!ht]
  \centering
  \includegraphics[width=1\linewidth]{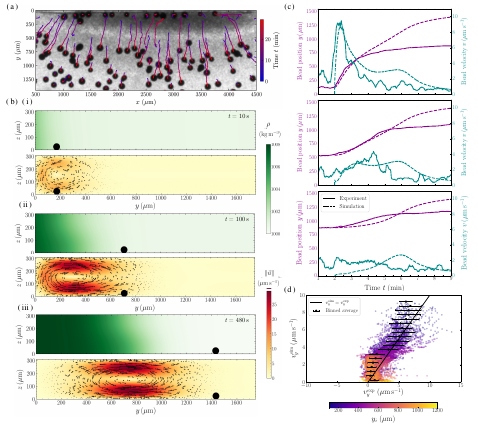}%
    \caption{
    \textbf{Bead dynamics within bioconvective flows.} 
    (a) Experimental trajectories of beads of diameter $d_b = \qty{110}{\um}$, in a chamber of height $H=\qty{310}{\um}$, pushed away from the chamber lateral wall by phototactic \chlamy\ accumulating at the top boundary, $\mathrm{OD}_i = 10$  (top view). 
    (b) Simulated evolution of the local density $\rho$ in the chamber, alongside the flow field it generates. Red regions correspond to zones of high velocity and illustrate the formation and outward propagation of bioconvective rolls. The successive positions of a simulated bead of diameter $d_b = \qty{50}{\um}$ sitting on the floor of the chamber is indicated by black disks. (i): $t=\qty{10}{\s}$. (ii): $t=\qty{100}{\s}$. (iii): $t=\qty{480}{\s}$. See \href{https://seminaris.polytechnique.fr/share/s/qbYKmyknoKSa9CG}{Supplementary~Movie~4}.
    (c) Position $y$ normal to the wall (magenta), and velocity $v$ (teal) of three representative beads of diameter $d_b = \qty{50}{\um}$, initially located at increasing distances from the wall. Experimental data are shown as solid lines while simulated data with a $t_\mathrm{lag}\simeq2\,$min are displayed as dashed lines. Top: A bead, initially close to the wall ($y_i=\qty{125}{\um}$), is rapidly accelerated by the algae-induced bioconvective flow, reaching a peak velocity of approximately $\qty{10}{\micro\metre\per\second}$, before decelerating and entering a steady regime at $\qtyrange{0.5}{1}{\micro\metre\per\second}$. Experiment and simulations are in excellent agreement.
    Middle: A bead starting farther away ($y_i=\qty{545}{\um}$) is accelerated more gradually and reaches a lower peak velocity. Experiment and simulations are in very good agreement.
    Bottom: A bead initially far from the wall ($y_i=\qty{880}{\um}$) moves at much lower speeds. Experiment and simulations are in good agreement (see main text for discussion). 
    (d) Instantaneous simulated bead velocity $v_y^{\mathrm{sim}}$ as a function of the instantaneous experimental velocity $v_y^{\mathrm{exp}}$, for beads initially located within \SI{1200}{\um} of the wall for an initial optical density $\rm{OD}_i=10$. Their initial position $y_i$ is color-coded from purple ($y_i=\SI{107}{\um}$) to yellow ($y_i=\SI{1200}{\um}$). The black line shows $v_y^{\mathrm{sim}} = v_y^{\mathrm{exp}}$.
    }
    \label{fig:fig2}
\end{figure*}

In the experiment, we observe that the beads start to move roughly two minutes after the light is turned on, whereas in the numerical model motion is instantaneous. Correcting for this lag $t_\mathrm{lag}$ not captured by the model, we find very good agreement between simulations and experimental results, with both the correct shape for the time evolution of the bead velocity and a correct order of magnitude, which depends on the initial bead position. The motion of beads is better predicted for beads initially closer to the lateral wall, that are initially taken up in the convection roll and are advected away from the forming dense algal region at speeds up to $\approx \qty{10}{\um\per\minute}$ (see Figure~\ref{fig:fig2}c, top). This phase of motion lasts $\approx \qty{1}{\minute}$, during which beads can travel up to $\approx \qty{500}{\um}$, see magenta line in Figure~\ref{fig:fig2}c, top, for $2\leq t \leq \qty{3}{\minute}$. Then, these beads exit the core of the convection roll. They continue to be pushed away from the dense algal region, albeit at a much slower characteristic speed, of order $\approx \qty{0.5}{\um\per\minute}$. In contrast, beads initially far away from the wall never enter the core of the convection roll. Their experimental and simulated velocities are smaller than $\approx \qty{4}{\um\per\minute}$, see Figure~\ref{fig:fig2}c middle and bottom. The simulations slightly overestimate these beads velocity, likely because we have neglected the bead friction with the wall. Nevertheless, these beads also exhibit a similar slow speed of order $\approx \qty{0.5}{\um\per\minute}$ at longer times, $t \geq \qty{8}{\minute}$, while pushed away from the roll.

Now taking into account all tracked beads, we plot in Figure~\ref{fig:fig2}d the experimental instantaneous velocity against its simulated value for each bead at each time. 
We find again a good agreement between simulations and experiments, notably for bead velocities $v_{y}^{\rm exp} \gtrsim \qty{3}{\um\per\s}$, corresponding to beads initially within $\qty{400}{\um}$ of the channel wall, see Figure~\ref{fig:fig2}d. Simulations overestimate bead velocities when experimental bead velocities are small ($v_{y}^{\rm exp} \lesssim \qty{3}{\um\per\s}$), which occurs for beads further away than $\qty{400}{\um}$ from the wall initially. In particular, for beads with experimental velocities around $v_{y}^{\rm exp} \approx \qty{1}{\um\per\s}$, the corresponding simulated velocities reach up to $v_{y}^{\rm sim} \approx \qty{4}{\um\per\s}$. This hints again at effects slowing down the beads experimentally, such as lubrication or friction. The study of these second-order effects is left for future work. All individual bead velocity profiles, simulated and experimental, are shown in Supplementary~Figure~10.

\subsection{Collective bead motion is driven by the dynamically evolving roll of bioconvection}

Armed with our understanding of the motion of single particles, we now consider the motion of a collection of beads. Experiments show that dense beads accumulate along a line, away from the dense algal region, see pictures in Figure~\ref{fig:fig1}f,g. This line will hereafter be referred to as the front of beads, see Supplementary~Figure~5 for details on its definition and detection. Its dynamics, which effectively corresponds to the collective motion of hundreds of individual particles, is studied and quantified to understand its origin. In particular, at long times $t\gtrsim\qty{10}{\minute}$, what sets the position of this front of beads?

\begin{figure*}[!b]
    \centering
    \includegraphics[width=1\linewidth]{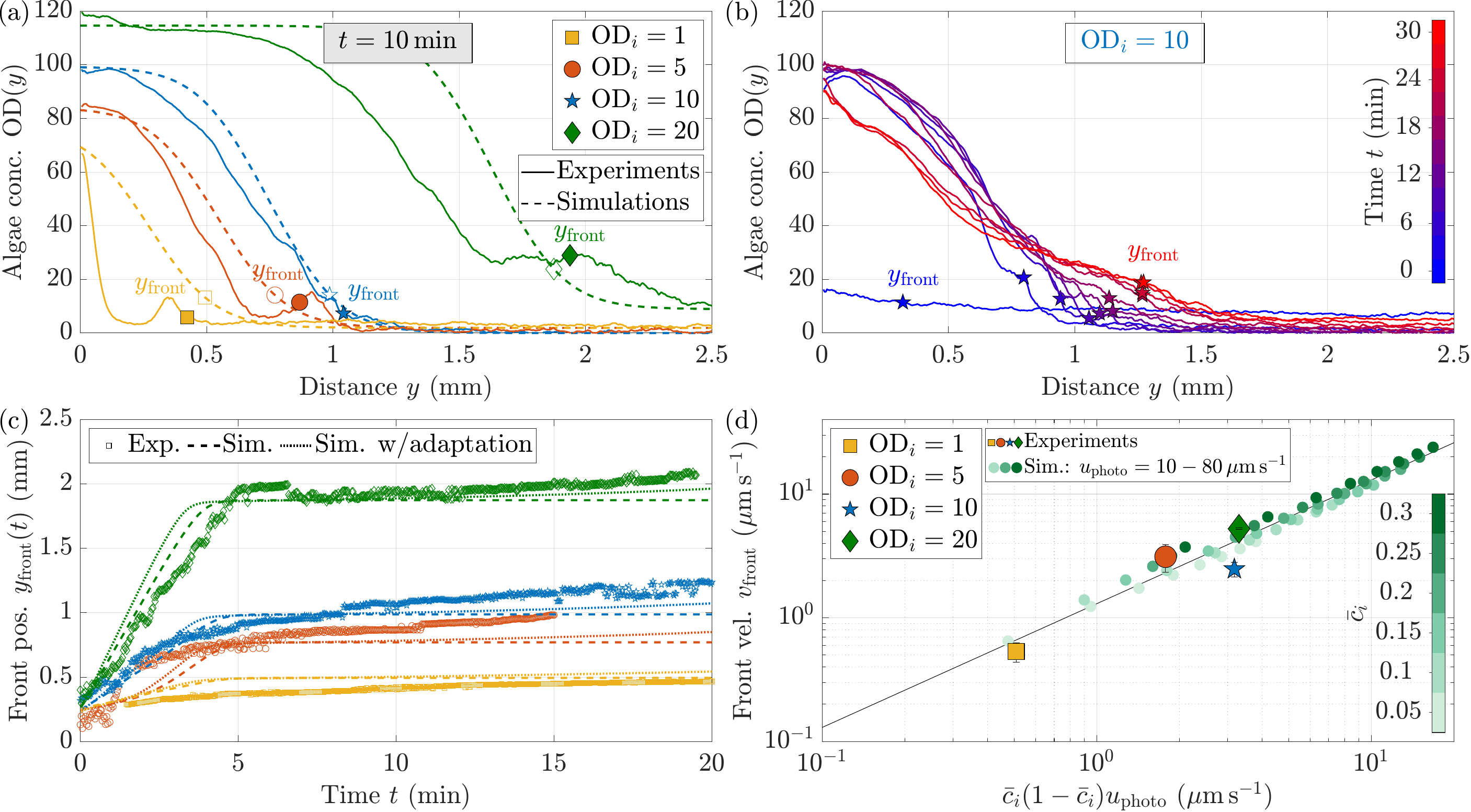}%
    \caption{\textbf{Denser beads form a front when repelled from dense algal regions.} (a) Profiles of algal concentration $\mathrm{OD}(y)$ near the lateral wall ($y=0$) for different initial optical densities OD$_i$. The profiles are taken $t=\qty{10}{\minute}$ after the blue LED is switched on. The position of the front of beads $y_\mathrm{front}$ is denoted for each curve with a filled colored symbol. Simulated profiles, in dashed lines, are also superimposed with their respective front position which is calculated as the position where the vertical fluid velocity is maximal (hollow symbols). (b) Temporal evolution of the experimental algal concentration profile $\mathrm{OD}(y)$ near the wall ($y=0$) for an initial optical density OD$_i=10$. Initially, there is no light stimulus so the profile is roughly flat. After the blue LED is switched on at $t=0$, the algae begin to accumulate at the wall. The position of the beads front $y_\mathrm{front}$ is denoted at each time with a colored star. See Supplementary~Figure~13 for a comparison between the experimental and simulated profiles. (c) Temporal evolution of the position of the beads front $y_\mathrm{front}$ for different initial optical densities OD$_i$, in the experiments (symbols) and in the numerical simulations without (dashed lines) or with adaptation (dotted lines). (d) Front velocity $v_\mathrm{front}$ scales linearly with $\uphoto \bar{c}_i(1-\bar{c}_i)$, both in the simulations (small symbols) and in the experiments (large symbols with errorbars). Simulations were done with a normalized initial concentration $\bar{c}_i=c_i/\cmax$ ranging from 0.05 to 0.3,  corresponding to $c_i=\qtyrange{1.1e7}{1.0e8}{\cells\per\mL}$ or OD$_i=\qtyrange{3.8}{34}{}$, and a phototactic velocity $u_\mathrm{photo}=\qtyrange{10}{80}{\um\per\s}$. Solid line: $v_\mathrm{front}=1.3\,\bar{c}_i (1-\bar{c}_i) u_{\rm photo}$.
    }
    \label{fig:fig3}
\end{figure*}

The initial cell seeding concentration is varied from $\qty{3e6}{\cells\per\milli\liter}$ to $\qty{6e7}{\cells\per\milli\liter}$, respectively corresponding to an optical density of 1 and 20, and the cells are subjected to the same directional light stimulus in all experiments. In all cases, algae accumulate away from the light source, triggering bioconvection. Dense $\qty{50}{\um}$ beads are then pushed away from the concentrated region in algae. The local algal concentration at all locations in the chamber is retrieved from the experimental images at all times, using the pixels' gray level intensity, see Supplementary~Figure~4. The algal concentration profile 10 minutes after the onset of stimulation is shown in \textbf{Figure~\ref{fig:fig3}}a as solid lines, for different initial cell seeding concentrations. As expected, the higher the initial concentration of algae, the larger the accumulation region. In each case, the position of the front of beads $y_\mathrm{front}$, superimposed on the concentration profiles as filled symbols in Figure~\ref{fig:fig3}a, is systematically several hundreds of microns away from the dense algal region.

For an initial optical density OD$_i=10$, the temporal evolution of the algal concentration profile is shown in Figure~\ref{fig:fig3}b. The near-uniform concentration profile at $t\leq0$, before the lighting of the blue LED, becomes peaked close to the lateral wall within minutes. Then, a stationary profile of algal concentration emerges, reflecting an equilibrium between the negative phototactic behavior of the algae, which makes them swim away from the light i.e. towards the wall at $y=0$, and both diffusion and convection, which tend to homogenize the concentration. At OD$_i=10$, the stationary state is reached in about 6\,min with the algae accumulation spanning from $y=0$ to 1\,mm. The front of beads, superimposed on the profiles as colored stars, is moving away from the dense concentrations appearing close to the wall, staying at the edge of the convection rolls in a region where the local concentration $\mathrm{OD}(y)$ does not exceed 20. For $t\geq\qty{10}{min}$, the concentration profile slowly flattens, resulting in a slower advance of the beads front after the initial burst of speed.

The evolution of the position of the front of beads $y_\mathrm{front}(t)$ with time is shown for four initial cell seeding concentrations, from OD$_i=1$ to OD$_i=20$, in Figure~\ref{fig:fig3}c. In all cases, the front displays an initial faster phase of motion during $\approx\qty{5}{\minute}$, with a quasi-constant speed of the order of a couple $\unit{\um\per\min}$, after which a slower front displacement is observed, with a speed of the order of $\approx \qty{0.1}{\um\per\min}$. These two phases of motion are all the clearer for higher values of the initial alga concentration OD$_i$.

Experiments are then compared to the numerical solution of Equation~\eqref{eq:NS1}-\eqref{eq:AD}, using as initial conditions a quiescent fluid and a concentration profile of algae identical to the experimental algae concentration profile at $t=0$ when the blue LED is switched on. A bioconvection roll forms in the simulation, and the front position in the simulations is defined as the position where the vertical fluid velocity is maximal, corresponding to the extremity of the convection roll. We find very good agreement up to $t\approx\qty{10}{\minute}$ between experiments and simulations, see the dashed lines and symbols in Figure~\ref{fig:fig3}a for the simulated concentration profiles and front position at $t=\qty{10}{\minute}$, respectively, and dashed lines in Figure~\ref{fig:fig3}c for the simulated front dynamics.

A front velocity $v_\mathrm{front}$ can be measured from the initial faster phase of motion of the front of beads ($t\leq\SI{5}{min}$). This velocity is quasi-constant, and can be simply defined as $v_\mathrm{front}^\mathrm{exp}=(y_\mathrm{front}(t)-y_\mathrm{front}(t_\mathrm{ref}))/(t-t_\mathrm{ref})$, for all values of $t$ such that $t_\mathrm{ref}<t\leq\SI{5}{min}$. The reference time $t_\mathrm{ref}$ is chosen as the first time the front can be reliably detected; we experimentally find $t_\mathrm{ref}=\SI{1.5}{min}$ for OD$_i=1$ and $t_\mathrm{ref}=0$ for all other initial concentrations. The front velocity is then calculated for $t=2$, 3 and \SI{4}{min}, with its average value being plotted in Figure~\ref{fig:fig3}d as large symbols and its standard deviation as error bars. To compare with numerical simulations, we assimilate $v_\mathrm{front}$ with the advance speed of the convection roll, which is also constant over the first phase of motion, see Figure~\ref{fig:fig3}c. 
Velocities from the simulations are computed for different normalized initial seeding concentrations $\bar{c}_i=\qtyrange{0.05}{0.3}{}$, corresponding to $c_i=\qtyrange{1.1e7}{1.0e8}{\cells\per\mL}$ or OD$_i=\qtyrange{3.8}{34}{}$, and different phototactic velocities $u_\mathrm{photo}=\qtyrange{10}{80}{\um\per\s}$. We find that the front velocity increases linearly with the product $\bar{c}_i (1-\bar{c}_i) u_{\rm photo}$, as highlighted by the small filled circles in Figure~\ref{fig:fig3}d. Experimental data on the front velocity are in excellent agreement with simulations, see large symbols in Figure~\ref{fig:fig3}d.

The front velocity $v_\mathrm{front}$ at early times ($t\lesssim\qty{5}{\minute}$) can be retrieved using the conservation of the number of algae. Indeed, the number of algae in the concentrated algal region is $ n(t) \approx \cmax H w(t)$, where $w(t)$ is the width of the concentrated region of algae at time $t$. The flux of algae into this region is due to algae from the bulk swimming towards the concentrated region: $dn/dt\approx c_i (1-c_i/\cmax) H u_{\rm photo}$, where $c_i$ is the algal concentration in the bulk. We therefore find 
\begin{equation}
    \frac{dw}{dt} \approx \frac{c_i}{\cmax} \left(1-\frac{c_i}{\cmax}\right) u_{\rm photo} = \bar{c}_i (1-\bar{c}_i) u_{\rm photo},
\end{equation}
with $\bar{c}_i=c_i/\cmax$, in excellent agreement with numerical simulations and experiments for $t\lesssim\qty{10}{\minute}$, see solid line in Figure~\ref{fig:fig3}d. 

Yet, longer time dynamics are not well captured by the model: the simulated front stops moving after $t\approx \qty{5}{\minute}$, see dashed lines in Figure~\ref{fig:fig3}c, while the experimental front of beads continues moving. This discrepancy can be understood by monitoring in experiments the time evolution of the algal concentration profile. As the algae concentration builds up at initial times ($t<\qty{6}{\minute}$), the front of beads moves over $\approx\qty{1}{\milli\meter}$. Then, after $t\approx\qty{12}{\minute}$ of light stimulation, the algae concentration profile flattens, see red lines in Figure~\ref{fig:fig3}b. While early times are well captured by Equation~\eqref{eq:NS1}-\eqref{eq:AD}, the flattening of the algae concentration profile, which is striking without beads clouding the concentration profiles (see Supplementary~Figure~12), is not reproduced by the model, as highlighted in Supplementary~Figure~13a-b. Indeed, this latter evolution is due to the adaptation of \chlamy\ to the light stimulus, a well-documented phenomenon.\textsuperscript{\cite{mayer1968chlamydomonas,arrieta2017phototaxis,harris2009chlamydomonas}} In response to a sustained strong light stimulus, algae initially experience negative phototaxis and swim away from the light; after a dozen minutes, they adapt to the light level and decrease their avoidance behavior, sometimes even switching to positive phototaxis after longer times, on the order of dozens of minutes to an hour.\textsuperscript{\cite{mayer1968chlamydomonas, arrieta2017phototaxis}} At the population level, adaptation corresponds to a mean phototactic velocity that continuously decreases, and can even change sign at long times.

To take into account adaptation in the model, the swimming speed is imposed to be a decreasing function of time, $u_\mathrm{photo} (t) = \left(u_\mathrm{photo}(0)-\alpha t \right)$, with $u_\mathrm{photo}(0)=\SI{43}{\um\per\s}$ and $\alpha=\SI{0.019}{\um\per\square\s}$. We then recover excellent agreement between simulated and experimental concentration profiles at all times (see Supplementary~Figure~13c), as well as between the simulated and experimental front dynamics, see dotted lines in Figure~\ref{fig:fig3}c. In particular, we retrieve the sustained slow motion of the front of beads at times larger than $t\gtrsim\qty{10}{\minute}$, due to the expansion of the convection roll.

\subsection{Collective particle motion: applications to surface cleaning and directional transport}

In configurations where beads are denser than the surrounding fluid and sediment, convective flows can be used to sweep the bottom surface of the environment. To highlight the potential of photo-bioconvective flows in surface cleaning applications, we define three quantitative performance metrics that we apply to the controlled experiments analyzed in Figure~\ref{fig:fig3}. The first metric, the bead surface fraction $\Psi_b$, quantifies how many beads are swept away from a predefined region of interest; the cleaning efficiency $\varepsilon$ measures how many beads are excluded from the dense algae region; last, the total cleaned surface $S_\mathrm{clean}$ assesses the extent of the region exempt of beads (``cleaned'') in the entire chamber.

\begin{figure*}[ht!]
  \centering
  \includegraphics[width=1\linewidth]{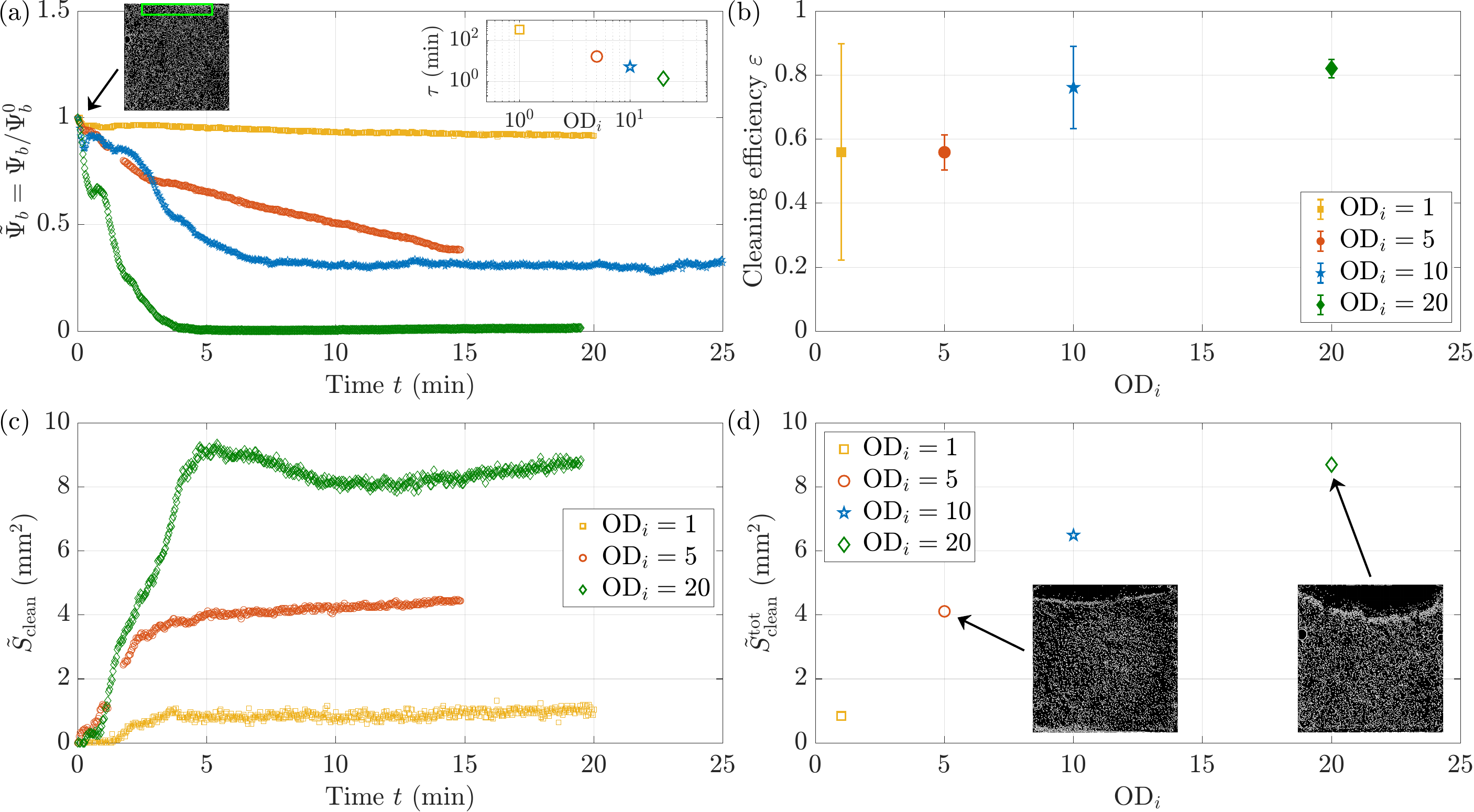}%
      \centering
    \caption{\textbf{Surface cleaning metrics.} (a) Temporal evolution of the bead surface fraction on the bottom surface $\Psi_b$, normalized by its initial value $\Psi_b^0=\Psi_b(t=0)$ before the blue light is switched on. $\Psi_b$ is measured in a \qtyproduct{6x0.9}{\mm} region, corresponding to \qtyproduct{67x10}{\%} of the whole chamber, located at the wall opposite of the light stimulus (see green rectangle in left inset, a typical image of beads in the entire \qtyproduct{9x9}{\mm} chamber at $t=0$). Right inset: Characteristic cleaning time $\tau$, calculated from the exponential fits $\Tilde{\Psi}_b=\Tilde{\Psi}_0\exp((t_0-t)/\tau)$. (b) Cleaning efficiency $\varepsilon$ of the beads by the algal bioconvection rolls, as a function of the initial optical density OD$_i$. $\varepsilon=1$ means that every bead was successfully swept away. $\varepsilon$ is measured by comparing the bead surface fraction, in the region located between the wall and \qty{200}{\um} ahead of the final bead front position, between $t=0$ and after the roll has vanished. (c) Temporal evolution of the cleaned surface $\tilde{S}_\mathrm{clean}(t)$ across the whole chamber. $\tilde{S}_\mathrm{clean}(t)$ corresponds to $S_\mathrm{clean}$ subtracted by the initial value $S_\mathrm{clean}^0=S_\mathrm{clean}(t=0)$ before the blue light is switched on. A region is considered cleaned if there are strictly less than 2 beads in a \qtyproduct{200x200}{\um} region. (d) Total cleaned surface $\tilde{S}_\mathrm{clean}^\mathrm{tot}$ across the entire chamber, as a function of the initial optical density OD$_i$. $\tilde{S}_\mathrm{clean}^\mathrm{tot}=\langle \tilde{S}_\mathrm{clean}(t)\rangle_{t=10..15\min}$. Insets: Snapshots of the beads in the entire \qtyproduct{9x9}{\mm} chamber at $t=20\,$min.
    }
    \label{fig:fig4_new}
\end{figure*}

First, the bead surface fraction $\Psi_b$ on the bottom surface is measured in a target region we aim to clean. This region is a rectangle of \qtyproduct{6x0.9}{\mm}, corresponding to \qtyproduct{67x10}{\%} of the entire chamber, located \SI{50}{\um} away from the wall opposite the light stimulus (see green rectangle in left inset of \textbf{Figure~\ref{fig:fig4_new}}a). Normalizing $\Psi_b$ by its initial value $\Psi_b^0$ at $t=0$, right before the blue light is switched on, $\Tilde{\Psi}_b=\Psi_b/\Psi_b^0$ decreases consistently for all experiments, indicating that beads are being swept away from the target region, see Figure~\ref{fig:fig4_new}a. The decrease is very small for the smallest initial optical density OD$_i=1$ ($c_i=\qty{3e6}{\cells\per\mL}$) with only 8\,\% of the beads being swept in 20\,min, whereas nearly all beads are moved away in this region of $\approx\qty{5}{\mm^2}$ for an initial optical density OD$_i=20$ ($c_i=\qty{6e7}{\cells\per\mL}$). The dynamics of cleaning likewise depend on the initial optical density. To estimate the characteristic time needed to clean the zone, the normalized fraction of beads $\Tilde{\Psi}_b(t)$ is fitted to an exponential decay. The characteristic cleaning time $\tau$ decreases monotonically when the initial algal concentration increases, spanning from a couple of minutes at the highest concentrations to more than 300\,min at OD$_i=1$, see right inset of Figure~\ref{fig:fig4_new}a.

The bead surface fraction in a predefined region does not provide the full picture, as it does not account for the total impact of the bioconvective rolls. Thus, we introduce other metrics that take into account the whole extent of the bioconvection rolls effect on the passive beads. When a roll is formed, not all dense beads are pushed away in the experiments due to impeding phenomena such as surface friction on the channel floor, local stickiness, or variability in bead buoyancies.   We define the cleaning efficiency  $\varepsilon$ to quantify the efficiency of the bioconvection roll for cleaning by comparing, in the entire region where the algae accumulated, the number of beads before and after the passage of the roll. Notably, the difference in beads is measured between $t=0$ and after the roll has completely vanished, to both avoid false negatives (some beads can be hidden by the high algal density) and also take the roll dispersion into account. The efficiency $\varepsilon$ increases significantly with the initial seeding concentration, going from roughly 50\,\% to 80\,\% when the initial OD goes from OD$_i=1$ to OD$_i=20$, see Figure~\ref{fig:fig4_new}b. This difference can be explained by the changes occurring in the rolls when the algal concentration increases: the rolls become both larger and faster when the initial OD rises (see Supplementary~Figure~11). In particular, these lower flow velocities could be too small to induce a motion of all beads, due to their interactions with the chamber floor. Note also that, while beads in a batch exhibit a range of buoyancies, between $\approx 1.001$ and $\approx 1.009$ with a median density $\approx 1.006$, they are always denser than the initial solution of algae and sediment to the bottom of the chamber at the beginning of the experiment. They are then mostly swept away by the forming bioconvection roll before the local algal concentration reaches its maximum value. Beads remaining in the zone of the convection rolls could be those with a density smaller than the density of the local, highly concentrated algal region, or those that exhibit the most friction with the chamber floor.

As a last metric, the full cleaned surface $S_\mathrm{clean}$ is measured across the entire chamber. A surface is considered to be clean if there are strictly less than 2 beads in a \qtyproduct{200x200}{\um} region. Subtracting the initial value $S_\mathrm{clean}^0=S_\mathrm{clean}(t=0)$ before the blue light is switched on, we obtain $\tilde{S}_\mathrm{clean}(t)$ the curve shown in Figure~\ref{fig:fig4_new}c. 
The total surface cleaned by the algae $\tilde{S}_\mathrm{clean}^\mathrm{tot}$ is defined as the average value between 10 and 15\,min. As for previous metrics again, the total surface cleaned increases monotonically with the initial algal concentration, ranging from \qty{0.8}{\mm^2} at OD$_i=1$ to nearly \qty{9}{\mm^2} at OD$_i=20$, i.e. more than 10\,\% of the chamber, see Figure~\ref{fig:fig4_new}d. This difference can be explained by the fact that, at lower concentrations $c$, algae are not sufficiently numerous to form a thick dense region. As a consequence, the convection roll remains at the wall. At larger concentrations, the dense region grows until all algae have accumulated and the convection roll moves away from the wall, following the algae front (see \href{https://seminaris.polytechnique.fr/share/s/qbYKmyknoKSa9CG}{Supplementary~Movie~4}). The total zone swept by the convection roll is thus the width of the roll itself plus the width of the dense algae region.

\begin{figure*}[ht!]
  \centering
  \includegraphics[width=1\linewidth]{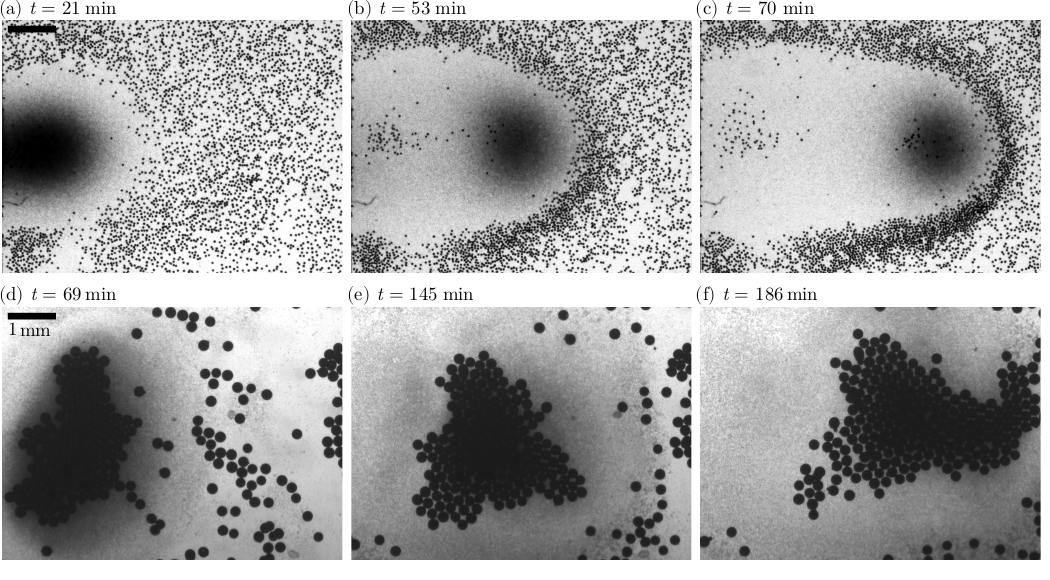}%
  
      \includegraphics[width=1\linewidth]{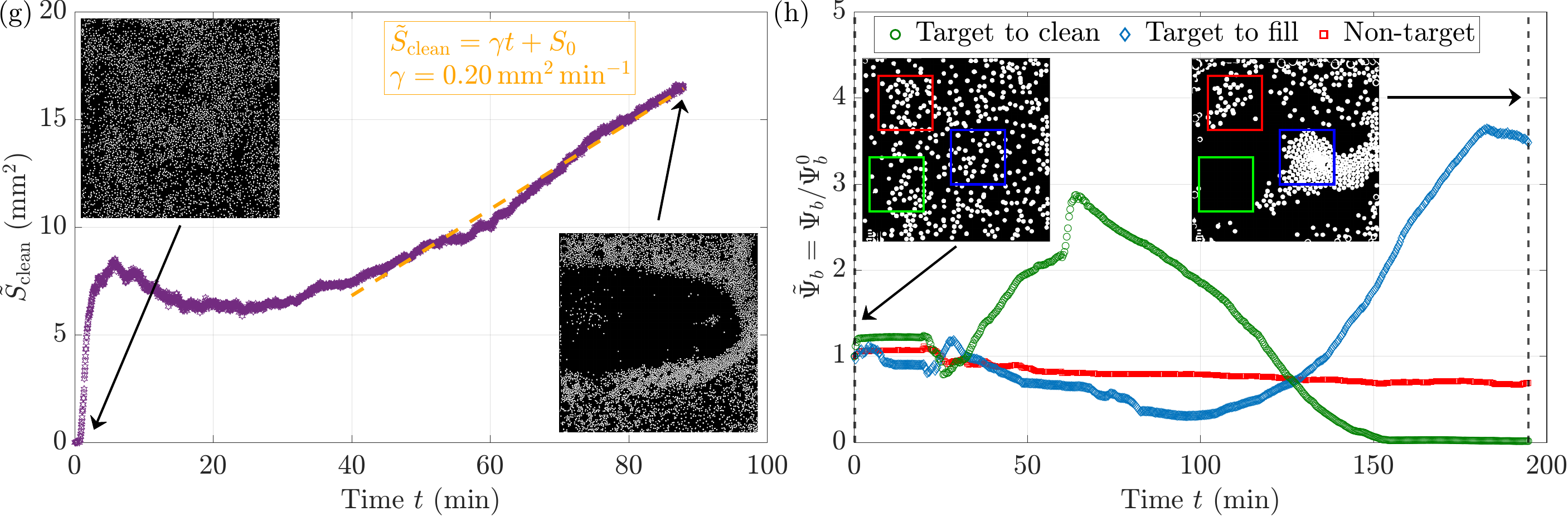}%
    \centering
    \caption{\textbf{Large-scale transport of large particles.} 
    (a)-(c) \textbf{An ``algae cannonball'' cleans the chamber.} Time-lapse of an experiment with \qty{50}{\um} beads in a dense \chlamy\ suspension at $\mathrm{OD}_i = 7$ (top view). Both lateral LEDs are switched on, creating a concentrated algal plume far from the walls. Denser beads are pushed away from the plume as it moves forward, see \href{https://seminaris.polytechnique.fr/share/s/BQz7jc2bp3dYFqm}{Supplementary~Movie~5}. In less than an hour, a region stretching over \qtyproduct{4x5}{\mm} is almost perfectly cleaned from beads. 
    (d)-(f) \textbf{Transporting a raft of beads.} Time-lapse of an experiment with \qty{230}{\um} beads in a dense \chlamy\ suspension at $\mathrm{OD}_i = 10$ (top view). By carefully adjusting LED intensities, the algal plume and bead raft are gradually moved, displacing the raft by almost 4\,mm over two hours. The latter collects neighboring beads as it moves, see \href{https://seminaris.polytechnique.fr/share/s/fqeBDiCN7Q8rsxY}{Supplementary~Movie~6}. 
    (g) Temporal evolution of the cleaned surface $\tilde{S}_\mathrm{clean}(t)=S_\mathrm{clean}(t)-S_\mathrm{clean}^0$ by the `algae cannonball' shown in (a)-(c). Orange dashed line: affine fit $\tilde{S}_\mathrm{clean}=\gamma t +S_0$ with $\gamma=\qty{0.2}{\square\mm\per\min}$ and $S_0=\qty{-1.2}{\square\mm}$. 
    (h) Temporal evolution of the bead surface fraction on the top surface $\Psi_b$ for the beads raft shown in (d)-(f), normalized by its initial value $\Psi_b^0=\Psi_b(t=0)$ before the blue light is switched on. $\Tilde{\Psi}_b$ is measured in three different regions of \qty{7}{\mm^2}: a target region to clean (green), a target region to transport the beads to (blue) and a control region (red) as reference. Their respective locations are shown in insets.}
    \label{fig:fig4}
\end{figure*}

To further bring the surface cleaning potential out, we place \qty{50}{\um} beads in a dense \chlamy\ suspension at $\mathrm{OD}_i = 7$ with both lateral LEDs on, forming a concentrated algal region away from the chamber walls. Axisymmetric convection rolls develop within a region of $\sim 1\,$mm radius, pushing beads outward within a larger influence zone of $\approx 2\,$mm, see Figure~\ref{fig:fig4}a and \href{https://seminaris.polytechnique.fr/share/s/BQz7jc2bp3dYFqm}{Supplementary~Movie~5}. By gradually adjusting the intensities of the LED bands, the algal plume is moved across the chamber, see Figure~\ref{fig:fig4}b. As the plume advances, it pushes denser beads away, effectively cleaning the chamber floor. Within an hour, an area of approximately \qtyproduct{4x5}{\mm} is nearly free of beads, see Figure~\ref{fig:fig4}c. The cleaned surface $S_\mathrm{clean}$ is further quantified as previously. Two regimes arise from its temporal evolution, see Figure~\ref{fig:fig4}g: at first, the cleaned surface quickly increases as the algal plume appears; then, it expands at a much slower rate, but consistently, as the bioconvection roll is directed towards the right side of the chamber. The cleaning rate observed on this curve is $\gamma=\qty{0.20}{\square\mm\per\min}$, a direct consequence of the approximately constant speed of the ``algae cannonball'' ($v_c\approx\qty{0.1}{\mm\per\min}$), and the approximately constant size of the exclusion zone ($d_c\approx\qty{2}{\mm}$), yielding a rate $v_c\cdot d_c \approx \qty{0.2}{\square\mm\per\min}$ identical to the measured rate. By the end of the experiment, after $\approx80\,$min, more than \qty{16}{\mm^2} are completely cleaned of beads. This performance is quite impressive, considering the size of the microorganisms involved is around \qty{10}{\um}.

Bead behavior in response to light-induced bioconvective flows depends on their relative density with the medium: lighter beads are attracted to high-concentration regions, while denser beads are pushed away. To illustrate the potential for directional transport in targeted delivery, we place buoyant \qty{230}{\um} beads in a dense \chlamy\ suspension at an initial optical density $\mathrm{OD}_i$ of 10. Both lateral LEDs were switched on with different intensities, creating a highly concentrated algae plume near the left wall, see \textbf{Figure~\ref{fig:fig4}}d and \href{https://seminaris.polytechnique.fr/share/s/fqeBDiCN7Q8rsxY}{Supplementary~Movie~6}. This plume exerts a strong attraction on floating beads, capturing them as the convection rolls develop and eventually forming a large aggregate, as can be seen in Supplementary~Figure~14. By gradually adjusting the LED intensities, with incremental steps spread over two hours, we move the algal plume across the chamber, pulling the raft of beads along its path, see Figure~\ref{fig:fig4}d-f. The raft grows by collecting beads along its path, reaching the opposite side of the chamber after two hours, having covered a distance of approximately 4\,mm and nearly doubling in size.

To quantify the performance of our system, we measure the bead surface fraction $\Psi_b$. Three different regions of area \qty{7}{\mm^2} are defined: a target region along the transport path, where beads should only travel transiently, a target region at the end of the transport path, where beads should end up accumulating, and a control region away from the transport path, which should not show any bead transport. These regions are shown in green, blue and red, respectively, in the inset in Fig.~\ref{fig:fig4}h. The normalized bead surface fractions $\Tilde{\Psi}_b$ are shown in Figure~\ref{fig:fig4}h for the three regions. After a transient step where  beads are temporarily gathered in the region to clean as the raft is formed, see Figure~\ref{fig:fig4}d, the raft is directed to its target, the blue region. As such, the normalized bead surface fraction in the region to clean (green) falls to 0 while it consistently increases to reach more than 3 in the target region to fill (blue). The number of beads in the control region (red) stays relatively put, only exhibiting a slight decrease as a few beads were attracted by the algal bioconvection roll. The consistent growth of the raft highlights the system’s ability to gather and transport passive objects within the active suspension, allowing us to direct the algal plume, and thus the bead raft, by precisely controlling LED intensities.

These observations and performance metrics demonstrate the potential for using algae-induced flows to manipulate passive objects in biological suspensions, paving the way for future research and industrial applications, such as micropollutant cleaning, demixing of impurities (see \href{https://seminaris.polytechnique.fr/share/s/dgkzSDY8BBFGm2z}{Supplementary~Movie~7} and \href{https://seminaris.polytechnique.fr/share/s/j9AnnzGtamqEbQx}{Supplementary~Movie~8} for simulations highlighting the demixing potential of the system) and particle sorting in microfluidic devices (see \href{https://seminaris.polytechnique.fr/share/s/n9E5PFb6E3tc6Wg}{Supplementary~Movie~9}).

\section{Discussion}

To summarize, by stimulating a suspension of \chlamyfull\ with light, we are able to transport hundreds of large passive particles over millimeters in a controlled manner. Transport is due to the formation of bioconvection rolls as algae, which are denser than water, accumulate horizontally in response to the light stimulus. Passive beads are either attracted or repelled by the rolls depending on their density with respect to the fluid, see Figure~\ref{fig:fig1}. The collective transport of beads is here mediated by the convective fluid flow, and not by steric interactions between individual algae and beads as in previous reports~\cite{jeanneret2016entrainment,leptos2009dynamics,ortlieb2019statistics,kurtuldu2011enhancement}. We show that a minimal numerical continuous model of phototactic bioconvection quantitatively captures most of the experimental observations, see Figure~\ref{fig:fig2}-~\ref{fig:fig3}. The bioconvection rolls are located at the edges of the concentrated algal region and extend over the area where there is a gradient in cell concentration, see Figure~\ref{fig:fig2}b. This concentrated region grows over time following simple scalings (see Figure~\ref{fig:fig3}d), and leads to up to 80\,\% of dense beads being swept away from the zone of bioconvection, with a clean zone of up to 8~mm$^2$, see Figure~\ref{fig:fig4_new}. The zone concentrated in algae can be dynamically moved by tuning the light stimulus, which leads to beads being swept away or accumulated along the path of the moving blob of algae (see Figure~\ref{fig:fig4}, \href{https://seminaris.polytechnique.fr/share/s/BQz7jc2bp3dYFqm}{Supplementary~Movie~5} and \href{https://seminaris.polytechnique.fr/share/s/fqeBDiCN7Q8rsxY}{Supplementary~Movie~6}). The zone of attraction or repulsion of the beads is determined by the roll size, a few hundred micrometers with our experimental parameters. Three different quantitative metrics show that, the larger the algae concentration, the larger the surface cleaned, and the faster and more efficient the cleaning (Figure~\ref{fig:fig4_new}).

The setup described here builds upon previous photo-bioconvection experiments\textsuperscript{\cite{dervaux2017light,arrieta2019light,ramamonjy2022light}} to drastically improve its applicability for cargo transport. Creating a lateral gradient of cell concentration as in\textsuperscript{\cite{dervaux2017light}} instead of a vertical one suppresses the cell concentration threshold required in\textsuperscript{\cite{arrieta2019light}} to trigger bioconvection. The use of phototaxis in the blue spectrum of light enables us to significantly reduce the time needed to establish bioconvection from approximately an hour in\textsuperscript{\cite{dervaux2017light}} to about a minute, as in.\textsuperscript{\cite{arrieta2019light}}. Note that the algae accumulation observed in our experiments stems from negative phototaxis. Positive phototaxis would lead to accumulation at the wall close to the light stimulus, and should also trigger the formation of bioconvection rolls. However, in this positive phototaxis configuration, at high cell concentrations, microalgae close to the wall may screen the light stimulus. Algae in the chamber would then be prevented from seeing the phototactic light, which would limit the size of the accumulated region, which is itself directly linked to the strength
of the bioconvection rolls. We therefore expect collective transport to be more efficient
when using negative phototaxis.

Harnessing macroscale bioconvective flows for microscale cargo transport is an appealing path. These flows enable the simultaneous displacement of hundreds of particles, each of which can be an order of magnitude larger than a single microorganism, and can either attract or repel particles depending on their relative density with the fluid. Besides, this directional transport can be dynamically tuned with external stimuli, e.g. light, controlling the microorganism `taxis'. These properties, absent from single-cell transport and multi-cell surface patterning methods, pave the way for more complex control strategies of microscale transport by microorganisms. Here, we demonstrate how to control the bioconvective flows of \chlamy\ suspensions with light, but the method is likely generalizable to other microorganisms (e.g. bacteria and other algae) and other external stimuli (e.g. chemicals\textsuperscript{\cite{raina2022chemotaxis}} and magnetic fields\textsuperscript{\cite{delong1993multiple}}). As long as the stimulus produces a lateral accumulation of microorganisms, similar sustained bioconvective flows will emerge.

Finally, in addition to cargo transport, these bioconvective flows can be used for both mixing and demixing. On the one hand, for colloidal particles or molecules, bioconvection enhances mixing. In Nature, this improves the distribution of nutrients and oxygen as well as light penetration.\textsuperscript{\cite{sepulveda2021persistence,di2023motile,shoup2023bacterial,zhu2025turbulent}} Controlling these bioconvective flows in biotechnological applications such as bioreactors could reduce operational costs and preserve cell viability by reducing mechanical steering and the associated shear stress felt by the cells.\textsuperscript{\cite{diaz1996mixing,arrieta2019light,carvajal2024towards}} On the other hand, bioconvection can be used to separate non-Brownian particles of different sizes and density. By attracting light beads and repelling dense ones with a size dependent interaction strength, bioconvection rolls have the potential to demix heterogeneous suspensions of particles.

\section{Experimental Section}

\threesubsection{Solution preparation}

\chlamyfull\ strain CC-125 (obtained from the Chlamydomonas Resource Center, University of Minnesota, MN, USA) was maintained by culturing on a solid medium consisting of Tris-Acetate-Phosphate (TAP, Gibco\texttrademark\ from ThermoFischer Scientific, France) and 1.5\% agar, with regular subculturing every 4 weeks. This cultivation method ensured the strain's motility, responsiveness to light, and minimized cell adhesion to surfaces. Liquid cultures were prepared every few days by inoculating algae from the solid culture into liquid TAP medium. These cultures were then incubated at 176\,rpm, under a 14-hour light/10-hour dark cycle, with a light intensity of \qty{60}{\umol\per\square\m\per\s}, and maintained at a constant temperature of \qty{22}{\celsius}. Maximum cell motility was typically reached within 3 days,\textsuperscript{\cite{harris2009chlamydomonas}} resulting in a suspension of actively swimming \chlamy\ cells at around \qtyrange{40}{60}{\um\per\s} and a diffusion coefficient $D\simeq\qty{4\pm1e-9}{\square\m\per\s}$ consistent with the literature\textsuperscript{\cite{polin2009chlamydomonas,arrieta2017phototaxis,dervaux2017light,arrieta2019light,fragkopoulos2025metabolic,wang2025light}} (see Section~I.C. in Supplementary).

Prior to experimental use, the liquid cultures were centrifuged to concentrate the algal suspensions, eliminate low-motility and dead cells, and remove cellular debris. In the first step, 45\,mL of the liquid culture was centrifuged at 1057\,g for 10 minutes. In the second, 39\,mL of the supernatant was discarded, and the remaining 6\,mL was centrifuged at 73\,g for 2 minutes. The resulting supernatant, containing the motile cells, was centrifuged once more at 285\,g for 5 minutes and the required amount of supernatant was removed to keep a solution at a specific concentration. Then, a TAP solution containing blue polyethylene microspheres (Cospheric, US), with diameters $d_b$ ranging from \qty{50}{\um} to \qty{460}{\um}, was added to the suspension of algae. The particle concentration was tuned to ensure a good repartition of the beads for imaging. Surfactant (Tween 20, Sigma Aldrich) was added to the solution at a concentration of 0.5\,\% to render the beads hydrophilic. We measured the densities of $\qty{50}{\micro\m}$ beads by placing them in solutions of salt of different concentrations. The beads were found to have a range of densities, between $\approx 1.001$ and $\approx 1.009$, with a median density $\approx 1.006$.

\threesubsection{Experimental setup}

Experiments were conducted within square chambers of width \qty{9}{\mm} and height $H=\qty{310}{\um}$ (Frame-Seal in situ PCR and Hybridization Slide Chambers, Bio-Rad), placed on top of a glass slide which was spin-coated with polydimethylsiloxane (PDMS, Dow-Corning Sylgard 184). The PDMS was made hydrophilic by plasma cleaning. Then, \qty{32}{\uL} of the algal solution, mixed with beads, was deposited within. The chamber was gently closed with a plasma-activated square PDMS lid. Experimental images were taken with a binocular (Leica MZ 16 FA) with a $1.0 \times$ objective (Plan APO) at 0.5\,fps. Since the algae do not react to red light, a red LED panel with wavelength $\lambda =\qty{630}{\nm}$ (TX Series Backlight, Metaphase lighting technologies) and a monochrome camera (IDS U3-3080CP-M-GL) were used for imaging. 
Two blue LED strips (Optonica, $\lambda =\qty{470}{\nm}$) of length 7.5\,cm were placed 3\,cm away from the assay, one on each side. We tested various experimental setups by adjusting the LED band’s distance from the chamber to achieve a light gradient predominantly in the $y$ direction while minimizing the transverse gradient, see Supplementary~Figure~2. The light intensity was tuned by varying the applied voltage and measured using a digital light sensor (Adafruit TSL2591). The experimental setup is sketched in Figure~\ref{fig:fig1}a,b. 
Initially, the blue-light is turned off and algae swim in all directions, populating the chamber homogeneously. In the experimental images, algae appear as a gray background, while the beads show up as large black dots, as shown in Figure~\ref{fig:fig1}d,h and Supplementary~Figure~1b. For large particles, the chamber height $H$ was increased with more chambers such that $d_b<H/2$ to allow free movement of beads of different sizes. To achieve an accurate measurement of algal concentrations, we correlated optical density (OD) measurements, which we use as a proxy, with the grayscale values obtained from 8-bit images, see Supplementary~Figure~3.

\threesubsection{Numerical simulations}

Numerical simulations of the continuum model, Equation~\eqref{eq:NS1}-\eqref{eq:AD}, are run in COMSOL Multiphysics v6.2. The numerical domain consists of a ($y$,$z$) slice of the experimental chamber of length $\qty{9}{\mm}$ and height $\qty{310}{\um}$. The Boussinesq Navier-Stokes equation is handled with the laminar flow module with P2+P1 elements and a volume force, while the active advection diffusion equation is handled with the general form PDE module with quadratic Lagrange elements. A triangular mesh with quadratic elements for the boundary layers is used. Mesh element sizes range between $\qty{5}{\um}$ and $\qty{18}{\um}$. At the chamber walls, we enforce the no-slip boundary condition for the fluid velocity and no-flux for the algae concentration $c$. In practice, we simulate the rescaled algae concentration $\bar{c}=c/\cmax$, which varies between 0 and 1, such that Equation~\eqref{eq:AD} reads 
\[\frac{\partial \bar{c}}{\partial t}+ \boldsymbol{\nabla}\cdot \left[(\mathbf{u}-u_\mathrm{photo} (1-\bar{c}) \mathbf{e_y})\bar{c} \right] = D\Delta \bar{c}.\]
The density dependence in Equation~\eqref{eq:NS1} with this rescaled concentration is then $\rho(\bar{c})=\rho_w+\bar{c}(\rho_\mathrm{max}-\rho_w)$ with $\rho_\mathrm{max}=\rho(c=\cmax)$. Equations are solved with the default fully coupled direct solver (MUMPS) and a relative tolerance of $10^{-3}$.

When comparing experiments and numerical simulation, we make sure to initiate the computation with the algae concentration profile observed experimentally just before the light is turned on. In experiments, we only have access to the algae concentration averaged over the chamber thickness $c(x,y,t)=\langle c(x,y,z,t) \rangle_z$. We thus use this average value as the initial condition for the concentration at every $z$: $c(y,z,0)=\langle c(y,0)\rangle$. We observe in experiments that the maximum concentration, and thus maximum density $\rho_\mathrm{max}$ increases slightly with the initial algae concentration in the range $\qty{1006}{\kilo\gram\per\cubic\m} <\rho_\mathrm{max}< \qty{1009}{\kilo\gram\per\cubic\m}$. The minimal optical density also does not always reach zero, indicating that some algae do not respond to light. We account for these non-motile algae by adjusting $\rho_w$ in the range $\qty{1000}{\kilo\gram\per\cubic\m} <\rho_w< \qty{1000.7}{\kilo\gram\per\cubic\m}$. As a first approximation, the value of $u_\mathrm{photo}=\qty{35}{\um\per\s}$ provides a good adjustment of the model for all initial algae concentrations.

To simulate the bead trajectories inside the photo-bioconvective flow, we assume the dense beads of diameter $d_b=\qty{50}{\um}$ to be passive tracers settled at the bottom of the chamber ($z=0$). At each numerical time step $\mathrm{d}t=0.5$\,s, each bead $j$ is advected at a velocity equal to the average horizontal fluid velocity over the bead diameter. Starting from an initial position $y_b^j(t=0)=y_i^j$ measured in experiments (see Figure~\ref{fig:fig2}a), we thus have $y_b^j(t+\mathrm{d}t)=y_b^j(t)+v^j(t) \mathrm{d}t$ with $v^j(t) =(1/d_b)\int_0^{d_b} \mathbf{u}(y_b^j(t),z,t).\mathbf{e_y} \mathrm{d}z$. 

The front speed is calculated from the position of the roll at 80\,\% of its maximum position reached in 20\,min: $v_\mathrm{front}^\mathrm{simu}=(0.8\,y_\mathrm{max}-y(0))/t(y=0.8\,y_\mathrm{max})$. This enables to take into account only the part of the data where the front speed is constant.

In \href{https://seminaris.polytechnique.fr/share/s/qbYKmyknoKSa9CG}{Supplementary~Movie~4}, we show the simulation reproducing the experiment with $\mathrm{OD}_i = 10$, including a simulated dense bead of $d_b=\qty{50}{\um}$. \href{https://seminaris.polytechnique.fr/share/s/dgkzSDY8BBFGm2z}{Supplementary~Movie~7} shows the simulation reproducing the experiment with $\mathrm{OD}_i = 20$, including several dense (black) and buoyant (white) beads of $d_b=\qty{50}{\um}$ randomly placed in the chamber to highlight the density based separation application. In \href{https://seminaris.polytechnique.fr/share/s/j9AnnzGtamqEbQx}{Supplementary~Movie~8}, we numerically reproduce the cannonball experiment by adjusting $\boldsymbol{u}_\mathrm{photo}$ as a function of time, to mimic the adjustment in light intensity from the two light bands in the experiment. \href{https://seminaris.polytechnique.fr/share/s/n9E5PFb6E3tc6Wg}{Supplementary~Movie~9} is a three dimensional generalization of Supplementary~Movie~8 illustrating particle transport in complex geometries. To simulate friction, a velocity threshold $u_\mathrm{tresh}=\qty{0.5}{\um\per\s}$ was included for the bead movement in Supplementary~Movies~7, 8 and 9. For more details, see Supplementary Material.

\medskip
\textbf{Supporting Information} \par 
Supporting Information is available from the Wiley Online Library or from the author.

\medskip
\textbf{Acknowledgements} \par 
We thank Caroline Frot for precious help in setting up the experiments and Victoria Nicolazo-Crach for fruitful discussions. This work was supported by ``Investissements d’Avenir'' LabEx PALM (ANR-10-LABX-0039- PALM) and by CNRS PEPS INSIS ``M\'{e}canique du futur''. J.B. was funded by the French Agence Nationale de la Recherche (ANR), under grant ANR-21-CE30-0044.

\medskip
\textbf{Data availability statement} \par
Data are available from the authors upon reasonable request.

\medskip
\textbf{Conflict of interest}  \par
The authors declare no conflict of interest.

\medskip
\textbf{Author contributions}  \par
T.L. and J.B. contributed equally to this work. J.B. and M.J. conceptualized the research. T.L, J.B., M.J. and G.A. designed the experiments. T.L. and J.B. performed the experiments and analyzed the data. E.J.-P. conceived and performed the numerical simulations. All authors discussed experimental and numerical results. G.A. and M.J. secured funding. T.L., J.B. and G.A. wrote the original draft of the manuscript. All authors reviewed the manuscript.

\medskip
\textbf{Materials \& Correspondence} \par
Correspondence and requests for materials should be addressed to Julien Bouvard and Mojtaba Jarrahi.

\medskip



\end{document}


\title{Supplementary Material -- Controlling the collective transport of large passive particles with suspensions of microorganisms}

\author{Taha Laroussi}
\thanks{These authors contributed equally to this work.}
\affiliation{\ladhyx}

\author{Julien Bouvard}
\thanks{These authors contributed equally to this work.}
\email[Contact author: ]{bouvard.julien@gmail.com}
\affiliation{\ladhyx}

\author{Etienne Jambon-Puillet}
\affiliation{\ladhyx}

\author{Mojtaba Jarrahi}%
\email[Contact author: ]{mojtaba.jarrahi@universite-paris-saclay.fr}
\affiliation{\fast}

\author{Gabriel Amselem}
\affiliation{\ladhyx}

\date{\today}
\maketitle

\section{Methods}
\subsection{Light intensity gradient within the chamber}

\begin{figure}[b]
    \centering
    \includegraphics[width=0.45\textwidth]{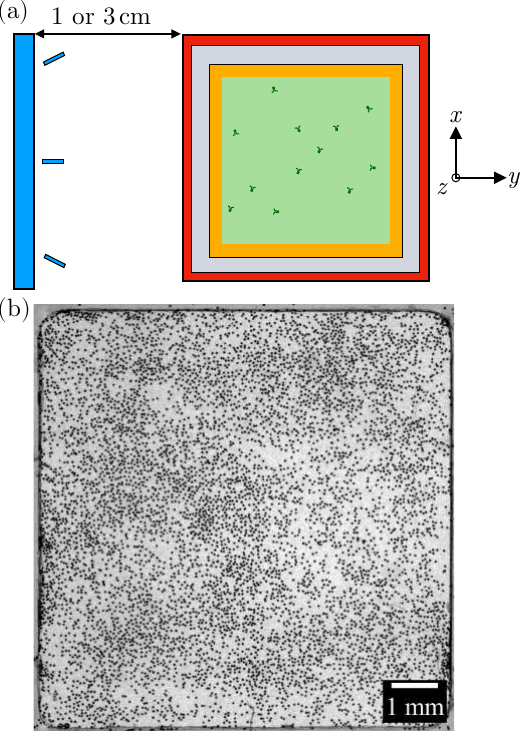}
    \caption{\textbf{Top view of the setup.} (a) Sketch of the \qtyproduct{9x9}{\mm} chamber filled with a suspension of \chlamyfull\ and passive polyethylene (PE) beads. The chamber is put on a PDMS-coated glass slide, on top of a red LED panel. (b) Top view of a chamber containing a suspension of \chlamy\ at an optical density OD$_i=7$ (\SI{2.1e7}{\cells\per\mL}) mixed with PE beads of diameter $d_b=\SI{50}{\um}$.
    }
    \label{fig:light_intensity_gradient_measure}
\end{figure}


Phototactic stimuli are provided by LED strips (LED SMD2835, Silamp). Each strip consists of five $2.8\times 3.5$~mm blue LED ($\lambda=\qty{470}{\nano\meter}$), equally spaced over 7.5\,cm, see Supp.~Figure~\ref{fig:light_intensity_gradient_measure}. The local light intensity within the chamber is estimated with a two-step process. First, the light intensity emitted by the LED strips is measured directly on both sides of the chamber using a digital light sensor (Adafruit TSL2591). Then, to measure the light intensity perceived by algae within  the chamber, the latter is filled with a solution of culture medium (Gibco\texttrademark~TAP, ThermoFisher Scientific, France) supplemented with 5\,\% fluorescein. When the blue LEDs are turned on, fluorescein is re-emitting light with a peak at 520\,nm, at an intensity proportional to the received blue light stimulus. The fluorescence signal is recorded with a fluorescence stereomicroscope (Leica MZ 16 FA) with a $1.0 \times$ objective (Plan APO) at 1900\,ms exposure time. Pixel gray values are proportional to the emitted fluorescence, and are used as a proxy to measure the light intensity profile within the chamber. The absolute value of the light intensities within the chamber are then obtained by combining pixel gray values with the light intensities previously measured with the digital light sensor.

\begin{figure}[t]
  \centering
  \includegraphics[width=1\linewidth]{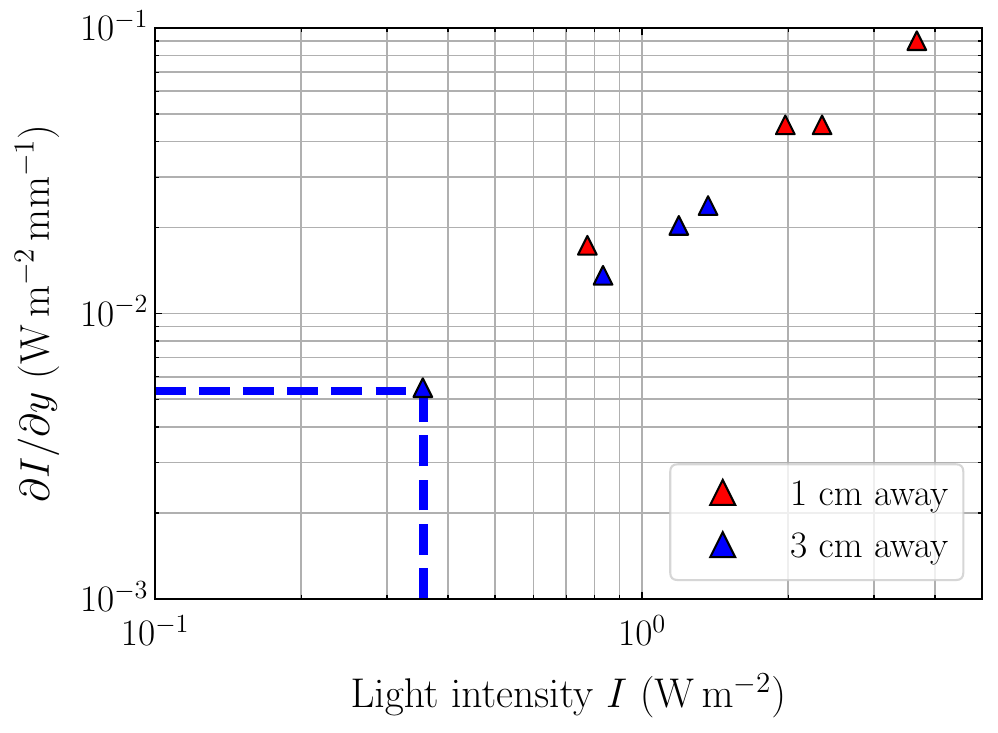}%
    \centering
    \caption{\textbf{Light intensity gradient within the chamber.} Light intensity gradient $\partial \mathrm{I} / \partial \mathrm{y}$ was measured for chambers positioned 1\,cm (red markers) and 3\,cm (blue markers) away from the blue LED band. Increasing the distance of the LED band to the chamber reduces the gradient. In our experiments, we place the chamber 3\,cm away from the LED band. Under these conditions, we find an experimental light intensity $I \approx \SI{0.35}{\W\per\square\m}$, represented by the dashed blue lines.}
    \label{fig:light_intensity_gradient_2}
\end{figure}

Different experimental setups were tested by varying the LED intensities and adjusting the LED strip's distance to the chamber, to achieve a light gradient predominantly in the $y$ direction while minimizing the transverse gradient in the $x$ direction. Our results indicate that both $\partial \mathrm{I} / \partial \mathrm{x}$ and $\partial \mathrm{I} / \partial \mathrm{y}$ across the chamber increases with higher LED intensities, see Supp.~Figure~\ref{fig:light_intensity_gradient_2} for the latter. Increasing the distance between the LED band and the chamber leads to a decrease in gradients, with a more pronounced reduction in $\partial \mathrm{I} / \partial \mathrm{x}$. Under identical experimental conditions, $\partial \mathrm{I} / \partial \mathrm{x}$ remains consistently lower than $\partial \mathrm{I} / \partial \mathrm{y}$: the gradient in $y$-direction is higher even without an optimal setup. Based on these results, we find that the optimal configuration involves positioning the LED band 3\,cm away from the chamber to minimize the light gradient in $x$-direction. This setup maintains the longitudinal gradient, which is crucial for our experiments. In our experiments, we thus place the chamber 3\,cm away from the LED band, with an applied voltage of 10\,V. Under these conditions, we find an experimental light intensity $I \approx \SI{0.35}{\W\per\square\m}$, represented by the dashed blue lines. These direct measurements allow for a more accurate assessment of the light intensity gradient compared to using gray values.

\subsection{Calibration of algal concentration measurements}
\begin{figure}[t]
  \centering
  \includegraphics[width=1\linewidth]{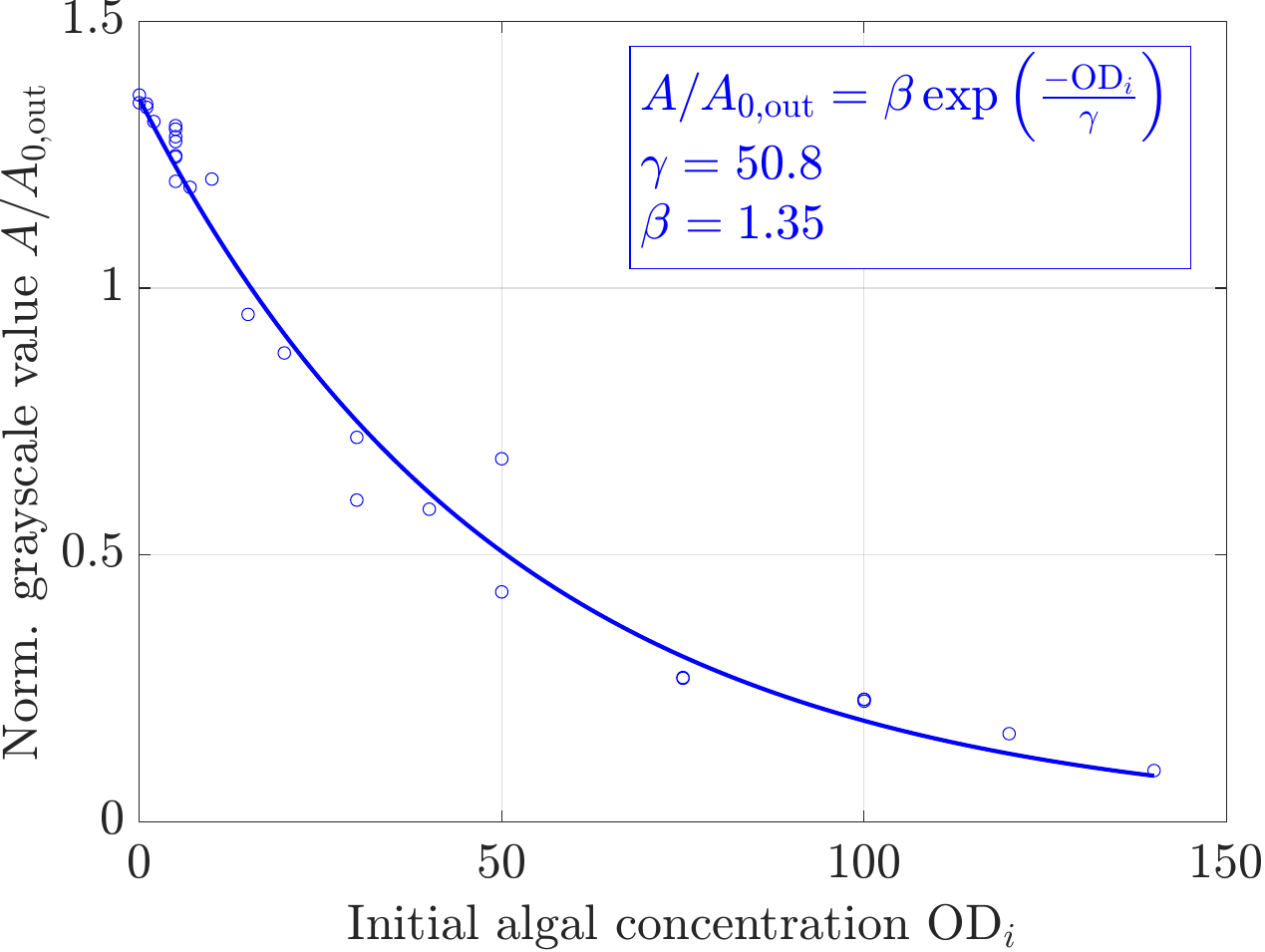}
    \centering
    \caption{\textbf{Calibration of algal concentration measurements.} Normalized grayscale values inside the chamber are plotted against the corresponding initial algal concentrations measured in optical density (OD). For reference, $1\,\mathrm{OD}=\SI{3e6}{\cells\per\mL}$. The concentration is measured with a spectrophotometer before filling the chamber with the algal solution. The blue line shows the best fit for the calibration.}
    \label{fig:Calibration_OD}
\end{figure}

To achieve an accurate measurement of algal concentrations, we correlated optical density (OD) measurements with grayscale values obtained from 8-bit images. For reference, $1\,\mathrm{OD}=\qty{3e6}{\cells\per\mL}$. We performed several calibration experiments, with varying algae concentrations, to map OD values to grayscale values. Before filling the chamber, the initial concentration of algae OD$_i$ was measured using a spectrophotometer. After filling, we captured an image of a large region where algae are distributed homogeneously and calculated the mean gray value within the chamber $A$, and outside the chamber $A_{\mathrm{0,out}}$. To account for differences in experimental conditions, such as variations in red LED panel light intensity or camera exposure time, the gray values within the chamber were normalized by those outside. The resulting data was fitted with a decreasing exponential, similar to Beer-Lambert law, in the form:
\begin{equation}
     A/A_{\rm{0,out}} = \beta\, \rm{exp}\left(-\frac{\rm{OD}_i}{\gamma}\right),
    \label{eq:Calibration_OD}
\end{equation}
with the best-fit parameters being $\gamma = 50.8$ and $\beta = 1.35$, see Supp.~Figure~\ref{fig:Calibration_OD}. The calibration curve derived from these experiments provides a mapping between grayscale values and the local algal concentration in OD.

\subsection{Algae motility}

\begin{figure}[ht]
  \centering
  \includegraphics[width=1\linewidth]{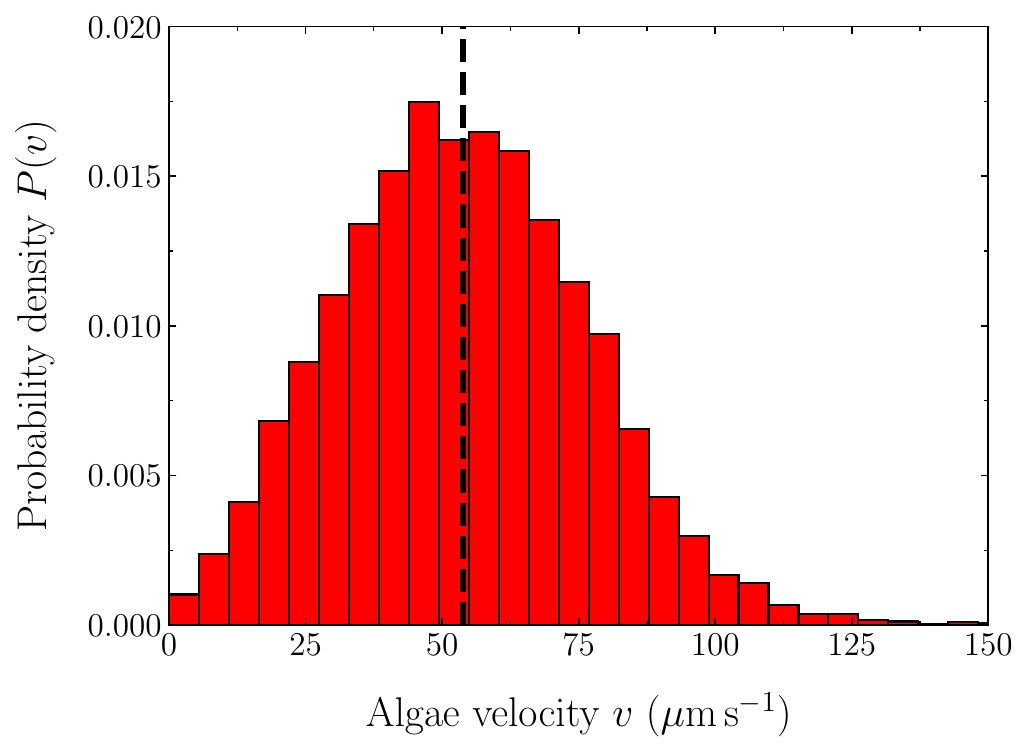}
    \centering
    \caption{\textbf{Typical distribution of the mean trajectory velocity} of \chlamy\ in 2D, for an experiment performed at $\mathrm{OD} = 1$ and tracked for 30\,s at 10\,fps. The average swimming speed, displayed in dashed black line, is \qty{54\pm23}{\um\per\s}.}
    \label{fig:uswim_histo}
\end{figure}

We characterize the motility of our algae \chlamyfull\ CC-125 by measuring their swimming speed alongside their diffusivity. Their swimming speed is measured by tracking individual \chlamy\ cells swimming in the bulk of a homogeneously filled closed chamber. Movies are recorded at \SI{10}{Hz} for \SI{30}{\s} under a Nikon TI microscope in a dark room, with a red filter (Newport RG645) to avoid any phototactic effects, using a $10\times$ objective and a CMOS camera (Hamamatsu ORCA-Flash4.0 LT, Hamamatsu Photonics, France). Tracking is performed in Fiji with the TrackMate plugin~\cite{tinevez2017trackmate}. A typical distribution of the mean algal velocity along its trajectory is shown in Supp.~Figure~\ref{fig:uswim_histo}, highlighting a large variation of the algae swimming speed peaked around $u_\mathrm{swim}=\qtyrange{40}{60}{\um\per\s}$.

To measure the diffusion coefficient of our algae \chlamy, we concentrate them in a small region of the chamber utilizing their negative phototactic response to intense blue light. Once accumulated on one wall of the chamber, we switch off the blue light stimulus. The algae thus diffuse away from the dense region, the temporal evolution of the algal concentration profile being consistent with a diffusive behavior. By fitting an isoconcentration with 
\begin{equation}
    y(t)=y_0 + \dfrac{2\sqrt{D}}{\mathrm{erf}(1-\mathrm{OD_{max}}/\mathrm{OD}_i)}\sqrt{t},
\end{equation}
we find the diffusion coefficient of the algae as $D=\SI{4\pm1e3}{\square\um\per\s}=\SI{4\pm1e-9}{\square\m\per\s}$. This value is consistent with previous diffusivity measurements of \chlamy~which found $D=\qtyrange{8e-10}{8e-8}{\square\m\per\s}$ depending on the strain and culturing conditions~\cite{polin2009chlamydomonas,arrieta2017phototaxis,dervaux2017light,arrieta2019light,fragkopoulos2025metabolic}, as well as the lighting conditions~\cite{wang2025light}.

\subsection{Bead front detection}

\begin{figure}[b]
  \centering
   \includegraphics[width=0.9\linewidth]{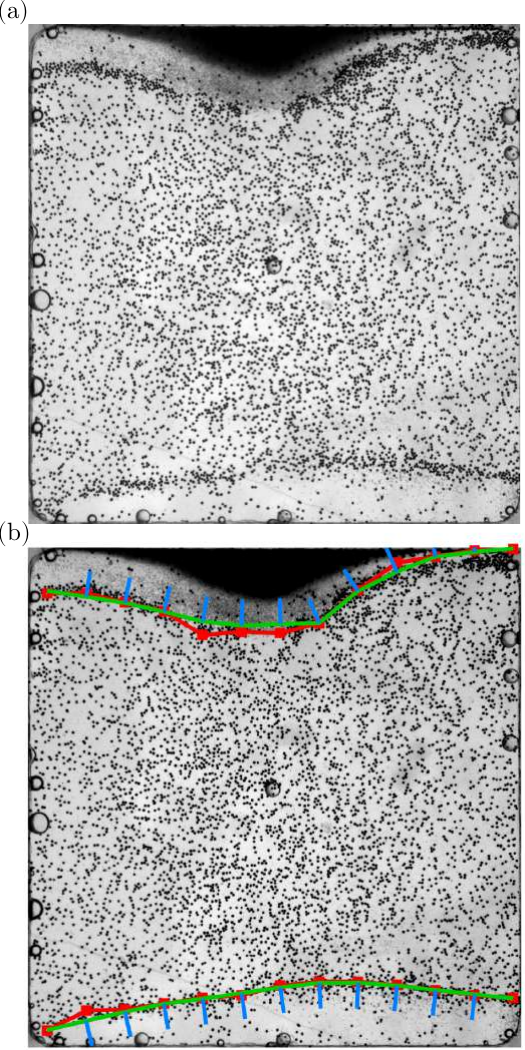}%
   \centering
    \caption{\textbf{Bead front detection}. (a) Image of the chamber with algae accumulated at the top boundary, pushing the beads away. Prior to this, the opposite LED was on, leading to the formation of another bead front near the bottom wall. The initial algae concentration is OD$_i = 5$. (b) Vertices of the bead fronts for each vertical slice, connected by straight lines, are shown in red. The smoothed front is shown in green. Normal directions to the smoothed fronts, at each vertex, are shown in blue, indicating the local direction of bead displacement.}
    \label{fig:supp_bead_front_detection}
\end{figure}

As algae accumulate, they tend to push the beads closer together, forming a dense line, referred to as the `front', see Supp.~Figure~\ref{fig:supp_bead_front_detection}a. Since the front may not be perfectly flat or parallel to the wall due to variations in the algae accumulation pattern, we divide the chamber into 13 vertical slices. This discretization of the front allows us to analyze each region independently. The bead distribution is then smoothed by applying a $20 \times 20$ uniform kernel to blur each slice. We then average the pixel intensity values along the $x$-direction within each blurred slice, producing a vertical profile of average gray values at each $y$-position. Since beads appear as dark spots in the images, the front is identified as the $y$-position where the pixel intensity reaches a local minimum, corresponding to the bead accumulation near the wall. These minimum positions represent the bead front for each slice and are shown as red vertices, see Supp.~Figure~\ref{fig:supp_bead_front_detection}b. The vertices from all slices are then connected with straight lines to reconstruct the continuous bead front. A locally weighted linear regression (``rlowess'') is applied to smooth the front, shown in green, see Supp.~Figure~\ref{fig:supp_bead_front_detection}b. The vectors normal to this smoothed front, shown in blue, indicate the local direction in which beads are pushed. Front detection can be difficult close to the wall or at early times, especially at low initial algae concentrations. In such cases, algae accumulation is thin and the wall can sometimes be misidentified as the bead front. To address this issue, we define dynamic detection boundaries that evolve with time: the boundary at each time step is adjusted to the previous position of the bead front $\pm\,20$ pixels, improving the accuracy of front detection over time. Additionally, each experiment undergoes visual verification of the bead front detection. Algae tend to accumulate more uniformly in the chamber's middle section. Unreliable vertices, especially those near corners, are often removed. Indeed chamber corners refract light and interfere with algae accumulation. This process allows for a robust detection of bead fronts and study of their dynamics.

\subsection{Numerical simulations - Supplementary Movies}

In \href{https://seminaris.polytechnique.fr/share/s/qbYKmyknoKSa9CG}{Supplementary~Movie~4}, we show the simulation reproducing the experiment with $\mathrm{OD}_i = 10$ ($\rho_w= \qty{1000}{\kilo\gram\per\cubic\m}$, $\rho_\mathrm{max}= \qty{1007.8}{\kilo\gram\per\cubic\m}$), including a simulated dense bead of $d_b=\qty{50}{\um}$.
\href{https://seminaris.polytechnique.fr/share/s/dgkzSDY8BBFGm2z}{Supplementary~Movie~7} shows the simulation reproducing the experiment with $\mathrm{OD}_i = 20$ ($\rho_w= \qty{1000.7}{\kilo\gram\per\cubic\m}$, $\rho_\mathrm{max}= \qty{1009}{\kilo\gram\per\cubic\m}$), including several dense (black) and buoyant (white) beads of $d_b=\qty{50}{\um}$ randomly placed in the chamber to highlight the density based separation application. 
In \href{https://seminaris.polytechnique.fr/share/s/j9AnnzGtamqEbQx}{Supplementary~Movie~8}, we numerically reproduce the cannonball experiment ($\rho_w= \qty{1000}{\kilo\gram\per\cubic\m}$, $\rho_\mathrm{max}= \qty{1009}{\kilo\gram\per\cubic\m}$, $\bar{c}_i=0.15$). To do so, $\boldsymbol{u}_\mathrm{photo}$ is adjusted as a function of time, mimicking the adjustment in light intensity from the two light bands in the experiment. In practice, the algae concentration field is made to follow a target location $y_{0}(t)$ that is time dependent such that $\boldsymbol{u}_\mathrm{photo}=-u_\mathrm{photo}(1-\bar{c}) \frac{(y-y_{0})}{\sqrt{(y-y_{0})^2}}\mathbf{e}_y$. In the movie, $y_{0}(t)$ is chosen as a piecewise function
\[y_{0}(t)=\begin{cases}
0 & \text{if } t \leq t_1,\\
v_c (t-t_1)  & \text{if } t \geq t_1.
\end{cases}\]
with $v_c=\qty{2}{\um\per\s}$ and $t_1=\qty{200}{\s}$. To simulate friction, a velocity threshold $u_\mathrm{tresh}=\qty{0.5}{\um\per\s}$ was included for the bead movement in Supplementary~Movies~7, 8 and 9.

\href{https://seminaris.polytechnique.fr/share/s/n9E5PFb6E3tc6Wg}{Supplementary~Movie~9} is a three dimensional generalization of Supplementary~Movie~8 illustrating particle transport in complex geometries. Equations~(1)-(3) from main text are solved on a three dimensional geometry representing a microfluidic chip with three wells. The mesh is coarser (element size $\sim\qty{50}{\um}$) and the parameters are $\rho_w= \qty{1000}{\kilo\gram\per\cubic\m}$, $\rho_\mathrm{max}= \qty{1009}{\kilo\gram\per\cubic\m}$, $\bar{c}_i=0.5$ in one of the well and zero elsewhere. The simulated blob is moved in 3D with a 2D target point $\{x_{0}(t),y_{0}(t)\}$, such that $\boldsymbol{u}_\mathrm{photo}=-u_\mathrm{photo}(1-\bar{c})\frac{(x-x_{0})\mathbf{e}_x+(y-y_{0})\mathbf{e}_y}{\sqrt{(x-x_{0})^2+(y-y_{0})^2}}$. This target point is moved from one chamber to another with a velocity $v_c=\qty{2}{\um\per\s}$. Such control of the algae could be achieved experimentally with a careful tuning of lights coming from six LED strips, two on each side of the three channels. 
Supplementary~Movie~9 shows a top view of the 3D simulation. The density is taken at the midplane, i.e. $\rho(z=H/2)$, and the velocities are taken at three quarters of the height, i.e. $\boldsymbol{u}(z=3H/4)$. The motion of buoyant beads of diameter $d_b=\qty{50}{\um}$ (white) randomly dispersed in two of the wells is simulated following the same procedure as in 2D, but for both the $x$ and $y$ components of the bead velocity.

\section{Results}
\subsection{Trajectories of beads close to convection rolls}

\begin{figure}[b]
  \centering
  \includegraphics[width=1\linewidth,trim=0mm 0mm 0mm 0mm,clip]{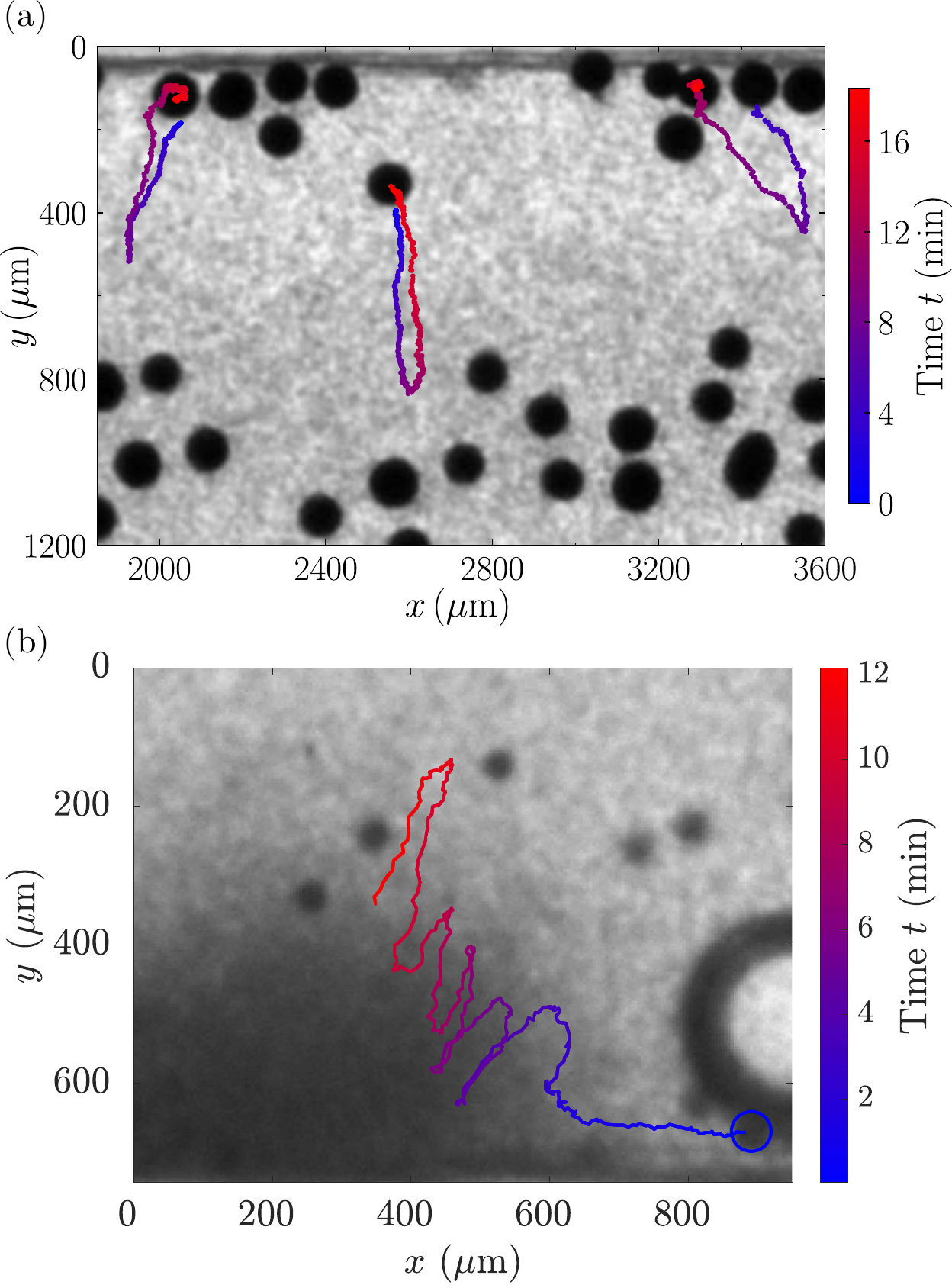}%
    \centering
    \caption{\textbf{Beads caught within convection rolls.} (a) Trajectories of three \SI{115}{\um} diameter beads caught within convection rolls. Initially, the beads are pushed away from the algae accumulation region at the wall. After traveling a few hundred microns, they are drawn back toward the wall. (b) Trajectory of a $d_b=\SI{50}{\um}$ bead trapped in a small convection roll. This bead exhibits multiple revolutions inside the roll, looking like it is `bouncing' on the edge of the dense algal region for nearly 1\,h, see \href{https://seminaris.polytechnique.fr/share/s/isWat6T4LsstFd6}{Supp.~Movie~3}. The initial algal concentration for both experiments is OD$_i = 10$.}
    \label{fig:repulsion_loop_trajectories}
\end{figure}

All experimental bead tracks are obtained with the TrackMate plugin~\cite{tinevez2017trackmate} in Fiji. As highlighted in the sketch of Main~Figure~1c, the bioconvection rolls created by algae accumulation cause the denser beads to be pushed away, while lighter beads are drawn into the convection rolls. The density of the beads determines whether they are repelled or drawn into the algae-induced flows.

Nevertheless, some beads closely match the density of the medium, allowing them to follow the flow patterns. At first, these beads are pushed away and move outward from the high algae density region, see Supp.~Figure~\ref{fig:repulsion_loop_trajectories}a. After traveling a few hundred micrometers, they enter upward-moving flows, which recirculate towards the wall, eventually drawing the beads back to the algae-dense region. This example highlights the simultaneous presence of inward (attractive) and outward (repulsive) flows at different heights within the chamber. As algae swim on average upward towards the wall, they displace the surrounding fluid, creating inward flows that transport beads. Once the algae accumulate, the unstable lateral concentration gradient causes the algae plume to sink, generating downwelling flows that push beads away from the high concentration zone. This cyclic movement is particularly visible for beads that closely match the density of the medium, as they experience minimal buoyancy, allowing them to follow the convection rolls and perform looping trajectories, see Supp.~Figure~\ref{fig:repulsion_loop_trajectories}b. Looking at this phenomenon from the top, it seems like the beads are `bouncing' on the edge of the dense algal region, see \href{https://seminaris.polytechnique.fr/share/s/isWat6T4LsstFd6}{Supp.~Movie~3}. For comparison between experimental and numerical results (Fig.~2d in the main text), all bead tracks starting at $t=0$ were considered, for beads that never entered upwelling flows.

\begin{figure}[ht]
    \centering
    \includegraphics[width=0.5\textwidth]{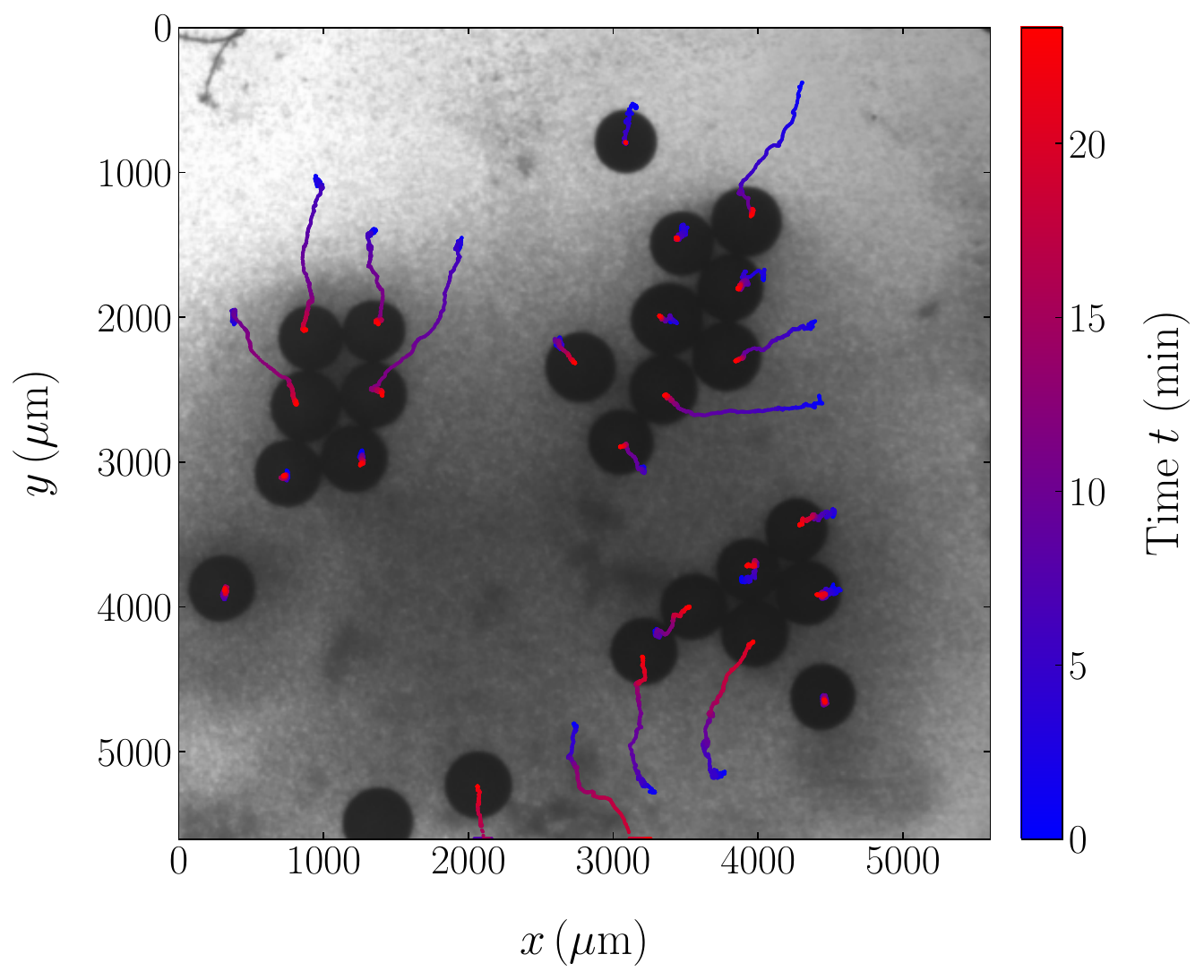} 
    \caption{\textbf{Trajectories of large beads attracted by plumes of algae.} Multiple algal plumes appear, locally attracting large floating beads of diameter $d_b = \SI{460}{\um}$ in their vicinity. Snapshots of the left aggregate constituted of 6 beads, are shown in Main~Figure~1f-k. See also \href{https://seminaris.polytechnique.fr/share/s/aQJpTfqWfPjMjm3}{Supp.~Movie~2}.}
    \label{fig:supp_460um_aggregates}
\end{figure}

Using lighter beads of diameter $d_b = \SI{460}{\um}$, we can show different bead dynamics in the vicinity of algal plumes. When a large algal plume appears, it attracts the floating beads nearby (see Main~Figure~1f-k and \href{https://seminaris.polytechnique.fr/share/s/aQJpTfqWfPjMjm3}{Supp.~Movie~2}). Trajectories of such beads are displayed in Supp.~Figure~\ref{fig:supp_460um_aggregates}.

\begin{figure}[b]
    \centering
    \includegraphics[width=0.5\textwidth]{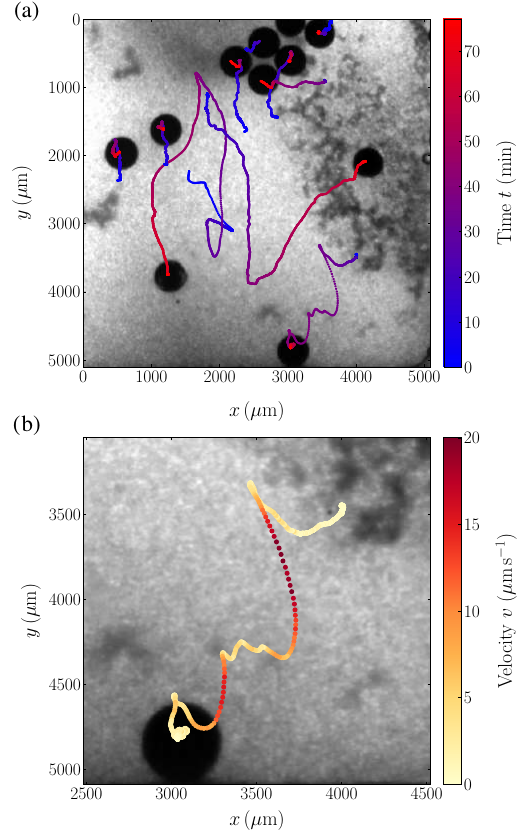} 
    \caption{\textbf{Long trajectories of large floating beads caught in multiple algal plumes.} Complex algae patterns appear over the course of the experiment, locally attracting buoyant beads of \SI{460}{\um} diameter in their vicinity. Depending on their location with respect to forming plumes, some beads can travel large distances (\SIrange{6}{9}{\mm}), hopping on and off as algal plumes develop and disappear. (a)  Several beads of diameter $d_b=\SI{460}{\um}$ exhibit complex, long-range trajectories as they go through different convection rolls over 77\,min. (b) A close-up view of a single large bead 'surfing' over forming algae plumes. This bead, of diameter $d_b=\SI{460}{\um}$, accelerates and decelerates twice as it goes through different convection rolls, reaching velocities as high as \SI{20}{\um\per\s} over 77\,min. Trajectories in (a) and (b) are superimposed on the last image of the trajectory taken at $t=77$\,min.}
    \label{fig:attraction_tracking_460um}
\end{figure}

\begin{figure*}[ht]
  \centering
  \includegraphics[width=1\linewidth]{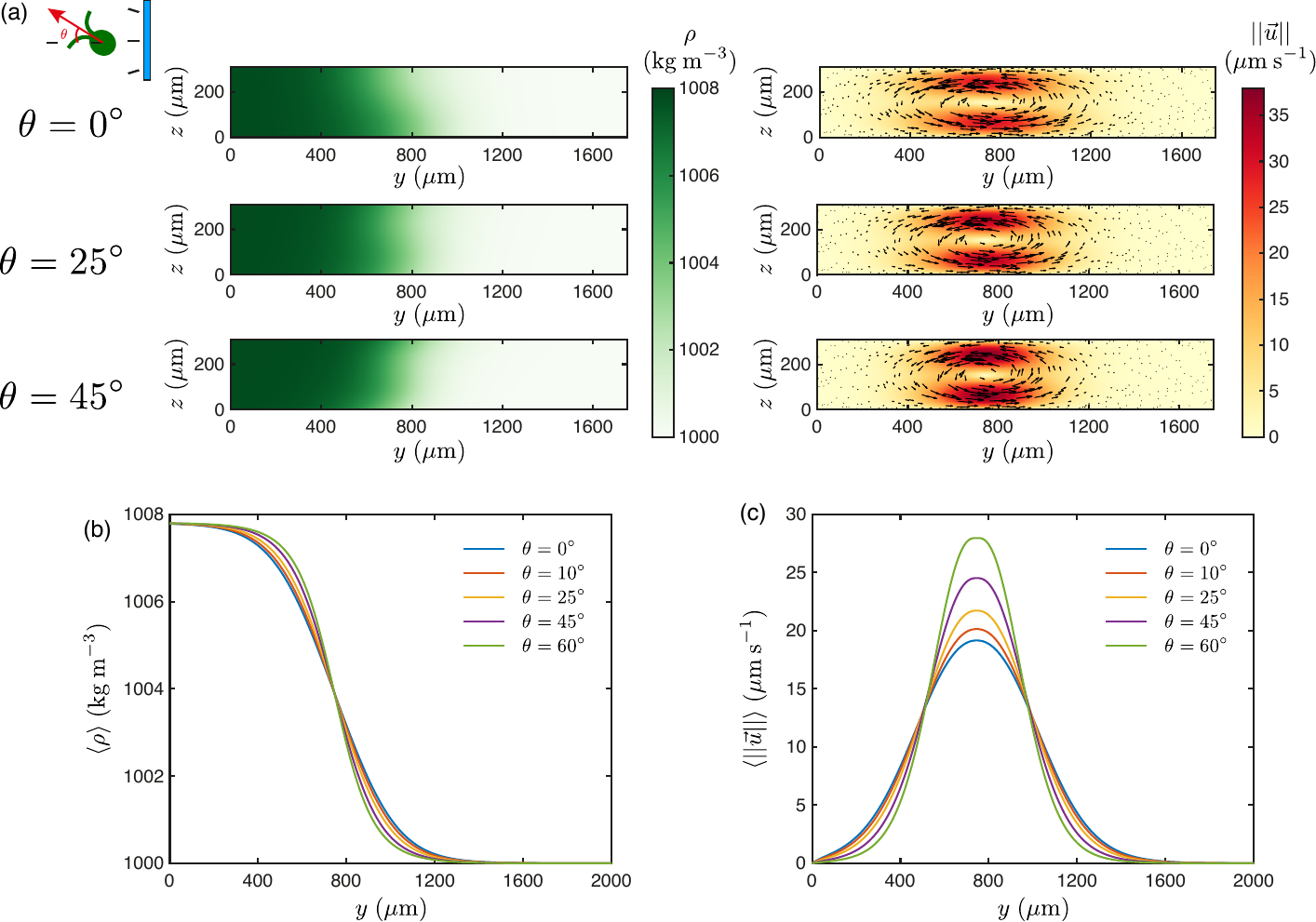}
    \centering
    \caption{\textbf{Effect of gravitaxis in the continuum model.} 
    (a) Density and velocity maps at $t=\qty{480}{\s}$ for simulations of the model with various values of the angle $\theta$ which introduces a gravitactic vertical velocity (see schematic on the top left and Eq.~\eqref{eq:gravitaxis}). The horizontal velocity  $\uphoto \cos\theta$ is kept constant for all simulations. (b,c) Depth averaged density $\langle\rho\rangle$ (b) and velocity magnitude $\langle||\mathbf{u}||\rangle$ (c) profiles at steady-state ($t=\qty{800}{\s}$) for different values of $\theta$.}
    \label{fig:sim_theta}
\end{figure*}

On a longer time scale ($\sim 1$\,h), some floating \SI{460}{\um} beads can 'surf' over the plumes, hopping on and off as they develop and disappear, allowing them to cover much longer distances that can reach \SIrange{6}{9}{\mm} over 80\,min, see Supp.~Figure~\ref{fig:attraction_tracking_460um}a. By zooming in on one of the beads, we see how it accelerates and decelerates twice as it goes through different convection rolls, reaching velocities as high as \SI{20}{\um\per\s}, see Supp.~Figure~\ref{fig:attraction_tracking_460um}b. This individual tracking approach highlights the complex response of beads to bioconvective flows, whether pushed away from algae-dense regions near the walls or drawn towards concentrated algal plumes.

\subsection{Numerical simulations}

In the minimal continuum model presented in Section~II.B. of the main text, we neglected gravitaxis which is known to affect \chlamy~\cite{bees2020advances}. We can include it in the model by changing the orientation of the swimming velocity $\mathbf{u}_{\rm photo}$. Assuming that the algae swim on average upward with an angle $\theta$ with respect to the light direction $\mathbf{e_y}$, the phototactic velocity reads
\begin{equation}
\mathbf{u}_{\rm photo}=\uphoto \cos\theta\left(1-\frac{c}{\cmax}\right) \left(-\mathbf{e_y}+\tan\theta\mathbf{e_z}\right).
\label{eq:gravitaxis}
\end{equation}
We solve Equations~(1)-(3) in main text using the phototactic velocity above in Equation~\eqref{eq:gravitaxis} for various values of $\theta$ in the conditions of Figure~2 of the main text, while keeping the horizontal component of the velocity constant, $\uphoto \cos\theta=\qty{35}{\um\per\s}$. The density and velocity maps for $\theta=\{\qty{0}{\degree},\qty{25}{\degree},\qty{45}{\degree}\}$ at $t=\qty{480}{\s}$ are shown in Supp.~Figure~\ref{fig:sim_theta}a. A higher value of $\theta$ tilts the density front and increases the magnitude of convection flows. However, the effects are small, even for large values of $\theta$ for which gravitaxis dominates phototaxis. The steady-state depth averaged density and velocity profiles are shown in Supp.~Figure~\ref{fig:sim_theta}b,c and confirms that the effect of $\theta$ is weak. Thus, gravitaxis is not critical to describe the phototactic bioconvection rolls observed in our experiment, and we discard it from the model presented in the main text.

The model introduces a crowding factor $\left(1-c/\cmax\right)$ in the swimming velocity $\mathbf{u}_{\rm photo}$ to restrict the algae concentration below $\cmax$. While several physical mechanisms could explain a decrease in swimming velocity at high algae concentration (steric repulsion, local cell stresses, light shielding, etc.), the choice of a linear dependence is arbitrary and was made for simplicity. Instead, the concentration dependency could be convex (i.e. strong at low concentrations and weak at high concentrations, like the light intensity in Beer-Lambert's law) or concave (i.e. weak at low concentrations and strong at high concentrations, like the viscosity of suspensions). We solved Equations~(1)-(3) of the main text with a crowding factor of the form $\left(1-c/\cmax\right)^n$ to test the effect of both a convex ($n>1$) and a concave ($n<1$) phototactic velocity concentration dependency. In practice, for $n<1$ the crowding factor was truncated at $c=0.995\cmax$ to avoid numerical difficulties, i.e. $\lim_{c\to\cmax} \left(\mathrm{d}\mathbf{u}_{\rm photo}/\mathrm{d} c\right)=\mathbf{\infty}$, and for $n>1$ $\cmax$ was slightly increased to reach the same density in the concentrated algal region. Qualitatively, the results are similar for all $n$: we always observe a density front and a convection roll. Quantitatively, the velocity magnitude decreases slightly with $n$ and the front/roll is sharper for $n<1$ and wider for $n>1$. When comparing the simulated height averaged density profiles to experiments (e.g. Figure~3a of main text), varying $n$ does not significantly improve the model, and therefore we keep $n=1$ for simplicity.

\begin{figure}[t]
    \centering
    \includegraphics[width=0.5\textwidth]{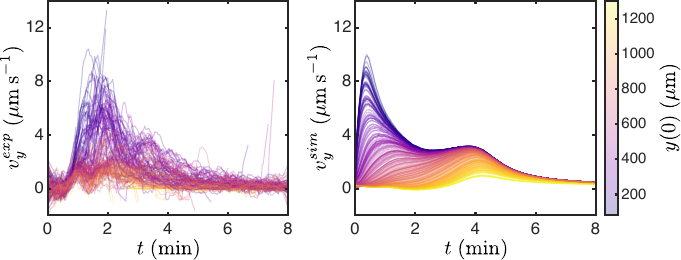} 
    \caption{\textbf{Experimental and simulated bead velocity profiles.}
    Experimental (a) and numerical (b) velocities of all the beads shown in Figure~2d of main text as a function of time (without lag). The color codes the bead initial position $y_i=y(0)$ with respect to the chamber wall.}
    \label{fig:velocityprof_expvssim}
\end{figure}

A side-by-side comparison of the experimental and numerical temporal velocity profiles of all the beads in Figure~2d of main text is shown in Supp.~Figure~\ref{fig:velocityprof_expvssim}. The numerical profiles are in good agreement with the experimental ones, exhibiting a similar shape and order of magnitude. The beads initially close to the wall ($y_i\leq\qty{500}{\um}$, in purple) are initially advected away by the convection roll at speeds up to $\approx\qty{10}{\um\per\minute}$ for $\approx1\,$min, before being slowly pushed away at $\approx\qty{0.5}{\um\per\minute}$. The beads initially far away from the wall ($y_i\geq\qty{900}{\um}$, in yellow) are displaced at velocities smaller than $\approx\qty{4}{\um\per\minute}$. Interestingly, even the two bumps that appear in the simulated profiles around $t\simeq30\,$s and $t\simeq4\,$min can be observed somewhat in the noisier experimental profiles.

\begin{figure}[ht]
    \centering
    \includegraphics[width=0.48\textwidth]{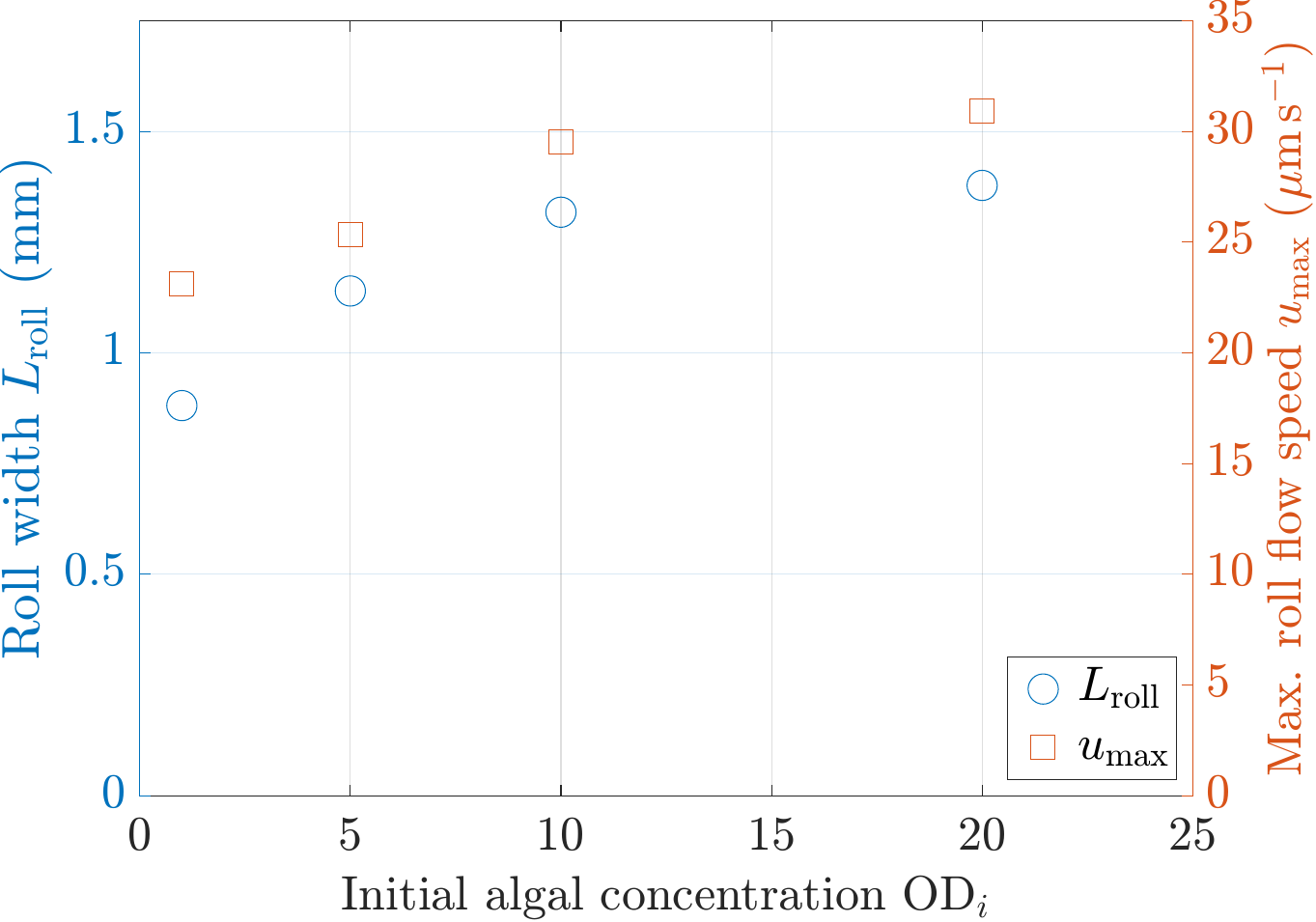} 
    \caption{\textbf{Characteristics of simulated bioconvection rolls.} 
    Width $L_\mathrm{roll}$ (blue) and maximum velocity $u_\mathrm{max}$ (orange) of the bioconvection roll, for different initial optical densities of algae, measured from the simulations. The roll width $L_\mathrm{roll}$ corresponds to the size of the roll in the $y$ direction (perpendicular to the wall), using a fluid velocity of \qty{0.5}{\um\per\s} as a threshold to define the roll edges.}
    \label{fig:simu_roll_size_vel}
\end{figure}

The numerical simulations allow for a characterization of the size and intensity of the bioconvection rolls when the initial seeding concentration of algae varies. The width of the roll $L_\mathrm{roll}$ and its maximum velocity $u_\mathrm{max}$ are measured in the simulations, see Supp.~Figure~\ref{fig:simu_roll_size_vel}, for the four algal concentrations used in the experiments ($\mathrm{OD}_i=1$, 5, 10 and 20). Both $L_\mathrm{roll}$ and $u_\mathrm{max}$ increase with the initial algal concentration, showing that the roll becomes both larger and stronger. Nevertheless, the variations are far from linear, with the roll size and the maximum velocity only increasing by respectively $\approx 35\,\%$ and $\approx 55\,\%$ when the optical density is increased 20-fold from OD$_i=1$ to OD$_i=20$.

\subsection{Algae adaptation to light}

Although the fast ballistic phase observed during $\approx 5\,$min can be readily attributed to the bioconvection rolls ejecting away the beads or attracting them, the existence of the last sustained slow regime in Figure~3c of main text, with a constant bead velocity $\approx \qty{0.5}{\um\per\min}$ is more complex to explain. We turn to monitoring of algal accumulation in the absence of beads, to better visualize the different phases of the algal response.

\begin{figure}[htb]
  \centering
  \includegraphics[width=1\linewidth]{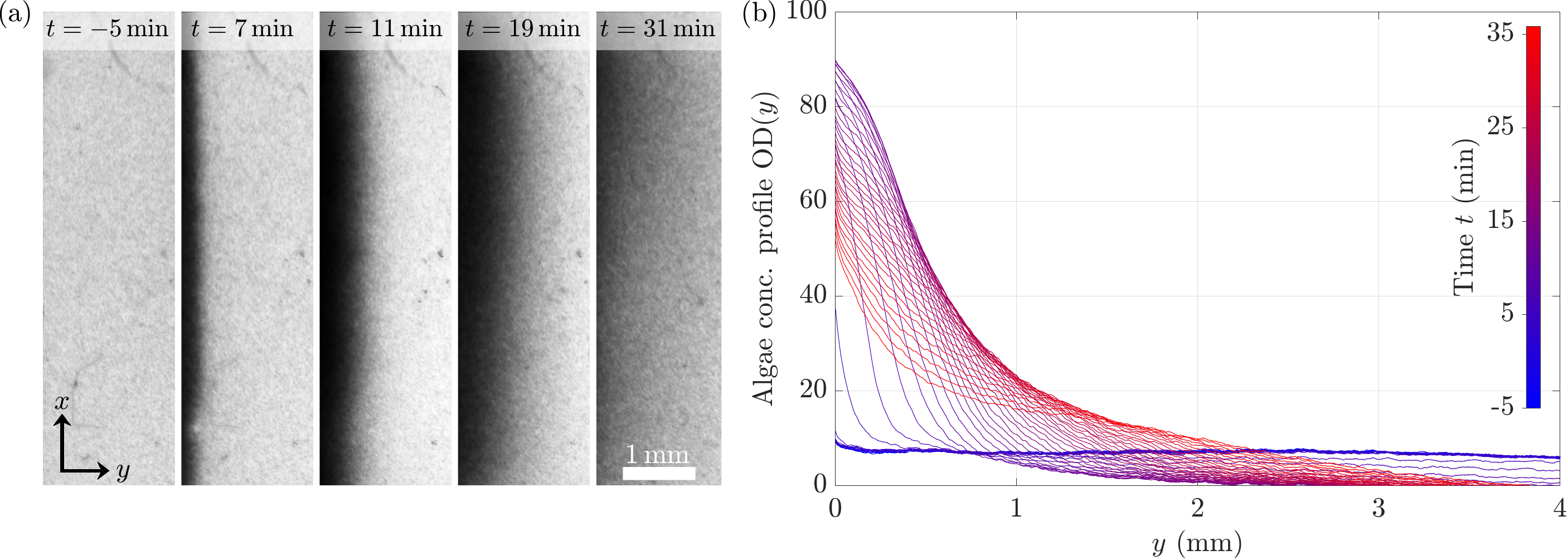}
    \centering
    \caption{\textbf{Phototactic adaptation to sustained light exposition.} (a) Time-lapse of an experiment without beads where a \chlamy\ suspension at $\mathrm{OD}_i = 10$ is exposed to a lateral light source coming from the right ($y>9\,$mm). Initially, algae are homogeneously distributed within the chamber. Once the lateral blue LED is switched on at $t = 0$\,min, algae exhibit negative phototaxis and rapidly accumulate at the opposite wall ($y=0$). By $t \approx 16$\,min, a dense region of algae has formed, spanning several hundred microns. A stationary state is reached when phototaxis is balanced by diffusion. Then, the phototactic response weakens due to adaptation which causes the algae to diffuse back into the rest of the chamber. (b) Temporal evolution of the algal concentration profile near the wall. At $t \leq 0$, the profile is flat as algae are uniformly distributed. A high concentration region forms close to the wall. Over time, the profile flattens as the phototactic response weakens due to adaptation, while still maintaining a high concentration region close to the wall.}
    \label{fig:light_adaptation}
\end{figure}

When a lateral light source is turned on, the algae accumulate at the opposite wall and concentrate in a dense band. The algal concentration in the band increases from $\mathrm{OD}_i = 10$ to as high as $\rm OD \approx 90$. The size of the band also increases with time, reaching $\approx \SI{500}{\um}$ after a dozen minutes of light stimulus, see Supp.~Figure~\ref{fig:light_adaptation}a. During the first dozen minutes, the band is very well defined, and we observe a sharp interface between the region dense in algae and the remaining of the chamber. After a dozen minutes, however, the initially sharp concentrated band both widens and blurs, an indication that algae start moving towards the light. After half an hour of light stimulus, the maximum algal concentration close to the wall is $\rm OD \approx 50$, but the band cannot be easily defined anymore. This blurring of the cell concentration can be attributed to cell adaptation to light, a well-known phenomenon~\cite{mayer1968chlamydomonas,arrieta2017phototaxis}. As cells adapt to light, their negative phototactic response dwindles, resulting in a decrease in their phototactic velocity $\uphoto$. Thus, the algal concentration profile flattens with time, lowering the concentration gradient, see Supp.~Figure~\ref{fig:light_adaptation}b.

The adaptation of algae to light is therefore responsible for both the sustained bead motion and the slowing down of their velocity. Indeed, in the absence of adaptation, the beads would be ejected away from the bioconvection rolls and remain immobile at the boundary of the rolls, as can be seen in \href{https://seminaris.polytechnique.fr/share/s/qbYKmyknoKSa9CG}{Supp.~Movie~4} for simulations without adaptation. Adaptation leads to a widening of the zone of high algal concentration, and so of the zone of bioconvection, which pushes the beads further away from the chamber walls. At the same time, the gradient of cell concentration decreases, and so does the lateral density gradient responsible for bioconvection rolls. This reduction in the density gradient leads to a slowing of the flow speed in the bioconvection rolls. 

\begin{figure}[b]
  \centering
  \includegraphics[width=1\linewidth]{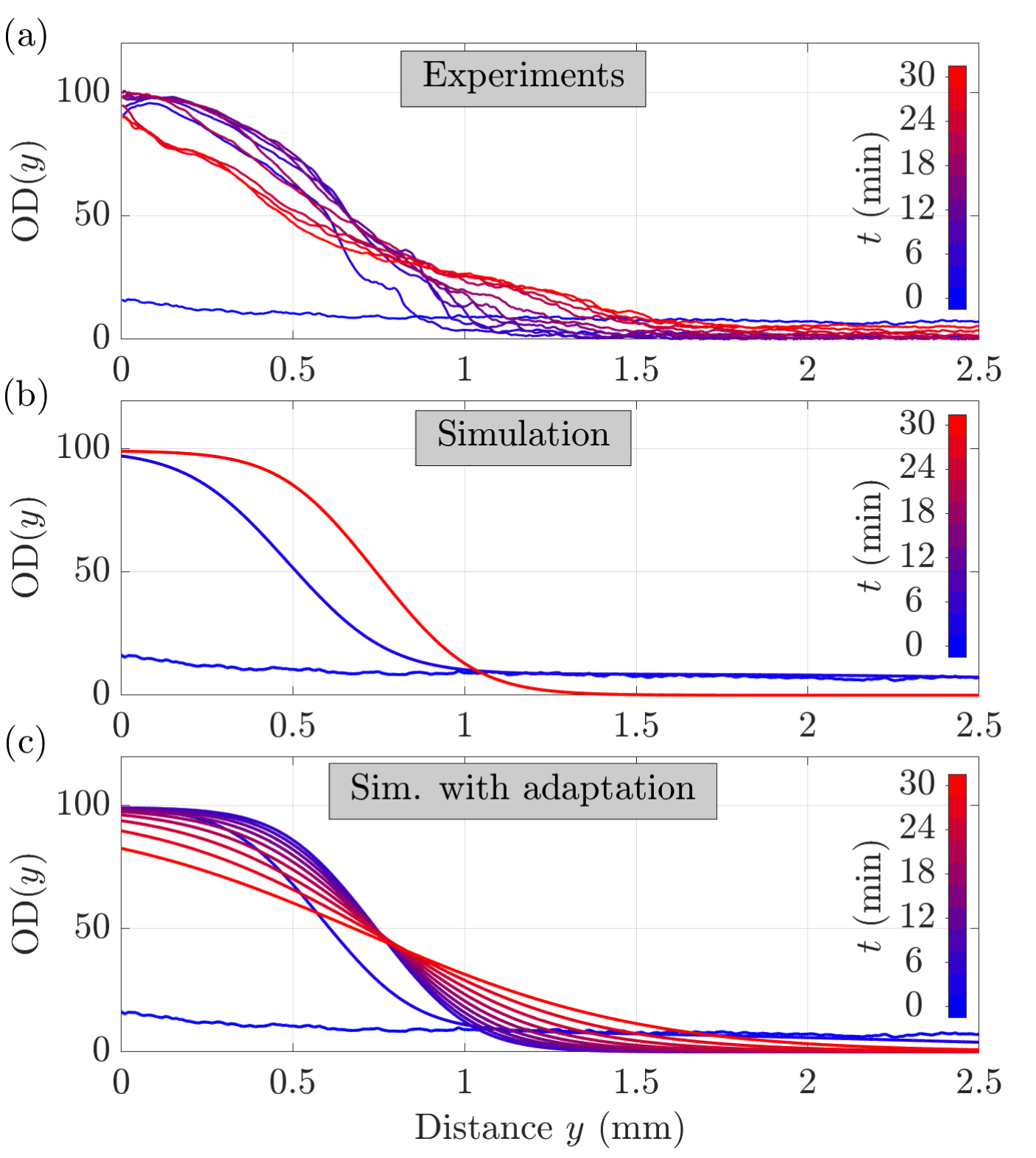}
    \centering
    \caption{\textbf{Adaptation is needed in the numerical simulations to match the experimental algal concentration profiles.} (a) Temporal evolution of the experimental algal concentration profile $\mathrm{OD}(y)$ near the wall ($y=0$) for an initial optical density OD$_i=10$. After the blue LED is switched on at $t=0$ and $y>9\,$mm, the algae begin to accumulate at the wall. (b) Temporal evolution of $\mathrm{OD}(y)$ obtained from the numerical simulations with a constant $\uphoto=\SI{35}{\um\per\s}$, i.e. without any light adaptation. (c) Temporal evolution of $\mathrm{OD}(y)$ obtained from the numerical simulations with light adaptation: $\uphoto(t)=\uphoto(0)-\alpha t$, with $\uphoto(0)=\SI{43}{\um\per\s}$ and $\alpha=\SI{0.019}{\um\per\square\s}$. The initial conditions for both simulations are a quiescent fluid and a concentration profile of algae identical to the experimental one at $t=0$ when the blue LED is switched on.}
    \label{fig:light_adaptation_simu}
\end{figure}

\begin{figure*}[ht!]
    \centering
    \includegraphics[width=1.0\textwidth]{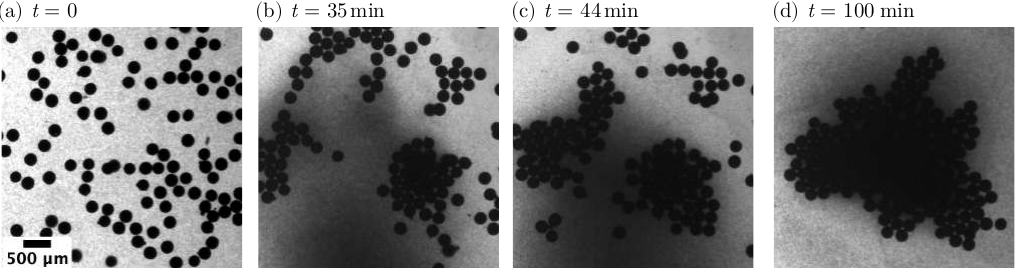} 
    \caption{\textbf{Raft formation}. Time-lapse of an experiment with $d_b=\SI{230}{\um}$ beads in a \chlamy\ suspension at OD$_i = 10$. Initially, both beads and algae are homogeneously distributed. Both side LEDs are switched on, creating a concentrated algae region far from the walls. Buoyant beads are drawn into the convection rolls and accumulate, forming a large raft of 158 beads around the plume. (a) $t = 0$. (b) $t = 35\,$min. (c) $t = 44\,$min. (d) $t = 100\,$min. See \href{https://seminaris.polytechnique.fr/share/s/fqeBDiCN7Q8rsxY}{Supp.~Movie~6}.}
    \label{fig:timelapse_raft_formation}
\end{figure*}

The numerical simulations based on algae swimming away from the light at a constant swimming speed $\uphoto$, i.e. without any light adaptation, do a good job in capturing the shape of the algal concentration profile when an equilibrium is reached, see Supp.~Figure~\ref{fig:light_adaptation_simu}. However, the slow time evolution at long times $t\geq10\,$min, which results in a flattening of the experimental profile, is not found in the simulations. By adding a time-dependent component to the swimming speed, such as $\uphoto(t)=\uphoto(0)-\alpha t$, we are able to properly capture both the initial accumulation of algae and its flattening at long times (see Supp.~Figure~\ref{fig:light_adaptation_simu}c).

\subsection{Raft formation}

To illustrate the potential for medium cleaning, we place floating \SI{230}{\um} beads in a dense \chlamy\ suspension at OD$_i = 10$. Both lateral LEDs were switched on with different intensities, creating a highly concentrated algae plume near the left wall, see Supp.~Figure~\ref{fig:timelapse_raft_formation}. This plume exerts a strong attraction on buoyant beads, typically those near the top lid. As convection rolls develop, the plume captures these lighter beads, eventually forming a large aggregate of more than 150 beads, over the span of an hour and a half, measuring around \SIrange{3}{4}{\mm} in length, see Supp.~Figure~\ref{fig:timelapse_raft_formation}d.

\newpage 

\bibliographystyle{apsrev4-2}
\bibliography{biblio}